%% file: paper.tex
\documentclass[11pt]{article}
\usepackage{graphicx}
\usepackage{amsmath}
\usepackage{epsfig}

\newcommand{\BABARPubYear}    {06}

\newcommand{\BABARConfNumber} {022}
\newcommand{\SLACPubNumber} {12014}

\input babarsym
\input mydef.tex

\long\def\inst#1{\par\nobreak\kern 4pt\nobreak
    {\it #1}\par\vskip 10pt plus 3pt minus 3pt}

\begin{document}
{\pagestyle{empty}
\begin{flushleft}
\end{flushleft}

\begin{flushright}
\babar-CONF-\BABARPubYear/\BABARConfNumber \\
SLAC-PUB-\SLACPubNumber \\
\end{flushright}

\par\vskip 3cm

\begin{center}
\Large \bf Determination of the Form Factors for the Decay {\boldmath \BztoDslnu} and of the CKM Matrix Element {\boldmath \Vcb}
\end{center}
\bigskip

\begin{center}
\large The \babar\ Collaboration\\
\mbox{ }\\
\today
\end{center}
\bigskip \bigskip

\begin{center}
\large \bf Abstract
\end{center}
\input abstract.tex
\vfill
\begin{center}

Submitted to the 33$^{\rm rd}$ International Conference on High-Energy Physics, ICHEP 06,\\
26 July---2 August 2006, Moscow, Russia.

\end{center}

\vspace{1.0cm}
\begin{center}
{\em Stanford Linear Accelerator Center, Stanford University, 
Stanford, CA 94309} \\ \vspace{0.1cm}\hrule\vspace{0.1cm}
Work supported in part by Department of Energy contract DE-AC03-76SF00515.
\end{center}

\newpage
} 

%
%
\input authors_ICHEP2006.tex

\input mypaper

\end{document}

%% file: mydef.tex
\newcommand{\bi}{\begin{itemize}}
\newcommand{\ei}{\end{itemize}}
\newcommand{\ben}{\begin{enumerate}}
\newcommand{\een}{\end{enumerate}}
\newcommand{\bc}{\begin{center}}
\newcommand{\ec}{\end{center}}
\newcommand{\bt}{\begin{table}}
\newcommand{\et}{\end{table}}
\newcommand{\be}{\begin{equation}}
\newcommand{\eeq}{\end{equation}}
\newcommand{\ba}{\begin{eqnarray}}
\newcommand{\ea}{\end{eqnarray}}

\newcommand{\la}{\ifmmode {\leftarrow} \else {$\leftarrow$}\fi}
\newcommand{\Ra}{\ifmmode {\Rightarrow} \else {$\Rightarrow$}\fi}
\newcommand{\La}{\ifmmode {\Leftarrow} \else {$\Leftarrow$}\fi}
\newcommand{\Lra}{\ifmmode {\Longrightarrow} \else {$\Longrightarrow$}\fi}
\newcommand{\Lla}{\ifmmode {\Longleftarrow} \else {$\Longleftarrow$\fi}}
\newcommand{\Llra}{\ifmmode {\Longleftrightarrow} \else {$\Longleftrightarrow$\fi}}
\newcommand{\Lk}{\ifmmode {{\cal L}} \else {${\cal L}$}\fi}
\newcommand{\Wt}{\ifmmode {{\cal W}} \else {${\cal W}$}\fi}
\newcommand{\Br}{\ifmmode {{\cal B}} \else {${\cal B}$}\fi}
\newcommand{\N}{\ifmmode {{\cal N}} \else {${\cal N}$}\fi}
\newcommand{\G}{\ifmmode {{\cal G}} \else {${\cal G}$}\fi}
\newcommand{\E}{\ifmmode {{\cal E}} \else {${\cal E}$}\fi}

\newcommand{\tBz}{\ifmmode {\tau_{\Bz}} \else {$\tau_{\Bz}$ }\fi }
\newcommand{\tBp}{\ifmmode {\tau_{\Bu}} \else {$\tau_{\Bu}$ }\fi }
\def\nubar    {\kern 0.18em\overline{\kern -0.18em \nu}{}\xspace}
\def\nulb     {\ensuremath{\nubar_\ell}\xspace}
\newcommand{\BtoDs}{\mbox{$\Bzb\rightarrow D^{*+} \ell^- \nulb$}}

\newcommand{\dsptoDz}{\mbox{$D^{*+}\rightarrow D^{0} \pi^+$}}
\newcommand{\psoft}{\ifmmode {\pi_s^+} \else {$\pi_s^+$}\fi }
\newcommand{\dm}{\ifmmode {\Delta M} \else {$\Delta M$}\fi}
\newcommand{\plab}{\ifmmode{p} \else {$p$}\fi}
\newcommand{\ks}{\ifmmode{k^*} \else {$k^*$}\fi}
\newcommand{\om}{\ifmmode{w} \else {$w$}\fi}
\newcommand{\omt} {\ifmmode {\tilde{w}} \else {$\tilde{w}$} \fi}
\newcommand{\mnusq}{\ifmmode{{M_\nu}^2} \else {${M_{\nu}}^2$}\fi} 
\newcommand{\DTau}{\ifmmode {\Delta \tau} \else {$\Delta \tau$}\fi}
\newcommand{\ggcc}{\ifmmode {GeV^2/c^4} \else {$GeV^2/c^4$}\fi}
\newcommand{\TBY}{\ifmmode{\theta_{\Bz, D^*\ell}} \else {$\theta_{\Bz, D^*\ell}$} \fi}
\newcommand{\Aone}{\ifmmode {{\cal A}_1} \else {${\cal A}_1$}\fi}
\newcommand{\Atwo}{\ifmmode {{\cal A}_2} \else {${\cal A}_2$}\fi}
\newcommand{\Rone}{\ifmmode {{\cal R}_1} \else {${\cal R}_1$}\fi}
\newcommand{\Rtwo}{\ifmmode {{\cal R}_2} \else {${\cal R}_2$}\fi}
\newcommand{\rha}{\ifmmode{\mbox{\rho^2_{{\cal A}_1}}} \else {\mbox{$\rho^2_{{\cal A}_1}$}}\fi}
\def\BpBm {\ensuremath{B^+ {\kern -0.16em \Bub}}}

\def\ctl        {\ensuremath{{\cos\theta_\ell}}\xspace}
\def\ctv        {\ensuremath{{\cos\theta_V}}\xspace}
\def\angchi     {\ensuremath{{\chi}}\xspace}

\def\thetal        {\ensuremath{{\theta_\ell}}\xspace}
\def\thetav        {\ensuremath{{\theta_V}}\xspace}

\def\ctl        {\ensuremath{{\cos\theta_\ell}}\xspace}
\def\ctv        {\ensuremath{{\cos\theta_V}}\xspace}

\newcommand{\Dsl}{\ensuremath{\Dstar\ell}\xspace}

\newcommand{\BztoDslnu}{\ensuremath{\Bz\rightarrow\Dstarm\ell^{+}\nul}\xspace}

\newcommand{\cBDl}{\ensuremath{\cos\theta_{\Bz,\Dstar\ell}}\xspace}

\newcommand{\KP}{\ensuremath{\kaon\pi}\xspace}
\newcommand{\KPPP}{\ensuremath{\kaon\pi\pi\pi}\xspace}
\newcommand{\KPPz}{\ensuremath{\kaon\pi\piz}\xspace}

\newcommand{\DzKP}{\ensuremath{\Dz\rightarrow\KP}\xspace}
\newcommand{\DzKPPP}{\ensuremath{\Dz\rightarrow\KPPP}\xspace}
\newcommand{\DzKPPz}{\ensuremath{\Dz\rightarrow\KPPz}\xspace}

%% file: abstract.tex
We present a combined measurement of the Cabibbo-Kobayashi-Maskawa
matrix element $|V_{cb}|$ and of the parameters $\rho^2$, $R_1$, and
$R_2$, which fully characterize the form factors of the $B^0 \rightarrow D^{*-}\ell^{+}\nu_\ell$
decay in the framework of HQET, based on a sample of about 52,800
$B^0 \rightarrow D^{*-}\ell^{+}\nu_\ell$ decays
recorded by the \mbox{\slshape B\kern-0.1em{\smaller A}\kern-0.1em
    B\kern-0.1em{\smaller A\kern-0.2em R}} detector. The kinematical information of the
fully reconstructed decay is used to extract the following values for
the parameters (where the first errors are statistical and the second
systematic):
$\rho^2 = 1.156 \pm 0.094 \pm 0.028$, $R_1 = 1.329 \pm 0.131 \pm
0.044$, $R_2 = 0.859 \pm 0.077 \pm 0.022$, $\mathcal{F}(1)|V_{cb}| = (35.03 \pm 0.39 \pm 1.15) \times 10^{-3}$.
By combining these measurements with the previous \mbox{\slshape B\kern-0.1em{\smaller A}\kern-0.1em
    B\kern-0.1em{\smaller A\kern-0.2em R}} measurements of 
the form factors which employs a different technique on a partial 
sample of the data, we improve the statistical 
accuracy of the measurement, obtaining:
$\rho^2 = 1.179 \pm 0.048 \pm 0.028,
R_1 = 1.417 \pm 0.061 \pm 0.044,
R_2 = 0.836 \pm 0.037 \pm 0.022, $ and $
\mathcal{F}(1)|V_{cb}| = (34.68 \pm 0.32 \pm 1.15) \times 10^{-3}.$
Using the lattice calculations for the axial form factor $\mathcal{F}(1)$,
we extract 
$|V_{cb}| =(37.74 \pm 0.35 \pm  1.25 \pm ^{1.23}_{1.44} ) \times 10^{-3}$, 
where the third error is due to the uncertainty in $\mathcal{F}(1)$.

%% file: authors_ICHEP2006.tex
\begin{center}
\small

The \babar\ Collaboration,
\bigskip

%
{B.~Aubert,}
{R.~Barate,}
{M.~Bona,}
{D.~Boutigny,}
{F.~Couderc,}
{Y.~Karyotakis,}
{J.~P.~Lees,}
{V.~Poireau,}
{V.~Tisserand,}
{A.~Zghiche}
\inst{Laboratoire de Physique des Particules, IN2P3/CNRS et Universit\'e de Savoie,
 F-74941 Annecy-Le-Vieux, France }
% ...

\end{center}\newpage

%% file: mypaper.tex
\section{INTRODUCTION}
\label{sec:intro}

The study of the \BztoDslnu decay~\cite{footnote1} is
interesting in many respects. In the Standard Model of electroweak
interactions the rate of this weak decay is proportional to the
absolute square of the Cabibbo-Kobayashi-Maskawa (CKM) matrix element
$V_{cb}$, which measures the weak coupling of the $b$ to the $c$
quark. Therefore the determination of the branching ratio of this
decay allows for a determination of $\Vcb$.

The decay is also influenced by strong interactions, whose effect can
be parameterized through two axial form factors \Aone\ and \Atwo, and
one vector form factor $V$, each of which depend on the momentum
transfer $q^2$ of the $B$ meson to the \Dstar\ meson. The form of this
dependence is not known {\it a priori}. Equivalently, instead of
$q^2$, the linearly related quantity $w$, which is the product of the
four-velocities of the \Bzb\ and \dsp, defined in Eq.~(\ref{eq:wdef}),
can be used.

In the heavy-quark effective field theory (HQET)
framework~\cite{ref:NeubertPhysReport,ref:Rich}, these three form
factors are related to each other through heavy quark symmetry (HQS),
but HQET allows for three free parameters which must be determined by
experiment.

Several experiments have measured \Vcb based on the study of the
differential decay width $d\Gamma/dw$ of \BztoDslnu
decays~\cite{ref:ARGUS,ref:BELLE,ref:ALEPH1,ref:ALEPH2,ref:OPAL,ref:DELPHI,ref:bad776}.
They determine only one of the parameters of the HQET from their data,
while they rely on the CLEO measurement of the other two, as described
in Ref.~\cite{ref:CLEOff}. This is the largest systematic uncertainty
common to all previous measurements of \Vcb based on this method.

Improved measurements of all parameters are also important for
describing the dominant background from $b\ra c\ra \ell$ cascades in
both inclusive and exclusive $b\ra u\ell\nu$ decays for measurements
of $|V_{ub}|$.  These motivations have been the basis of the
measurement of all three parameters performed by \babar\ and presented
in Ref.~\cite{ref:bad1224}.  While a new value of $\Vcb$ has been
computed using the new results as input to the previous measurement,
the full correlation between $\Vcb$ and the three form factor
parameters cannot be fully accounted for in that approach.

In the present analysis, we perform a simultaneous measurement of both
$\Vcb$ and of the form factor parameters from the measurement of the
full four-dimensional differential decay rate (see
Sec.~\ref{sec:formalism}).  We extend the method of
Ref.~\cite{ref:bad776} by combining the measurement of different
one-dimensional decay rates, fully accounting for the correlations
between them, to extract the whole set of observables.

A brief outline of the paper is as follows: In
Sec.~\ref{sec:formalism}, we introduce the definition of all used
observables, form factors, form factor combinations, and parameters.
In Sec.~\ref{sec:babar} we briefly describe the relevant aspects of
the detector and datasets we utilize in this measurement.  In
Sec.~\ref{sec:recoandsel} we describe event reconstruction and
selection and in Sec.~\ref{sec:analysis} the analysis method.
Sec.~\ref{sec:results} describes fit results and the estimation of
systematic uncertainties. In Sec.~\ref{sec:summary} we summarize the
results of this analysis and combine form factor measurements with a
previous \babar\ measurement to arrive at a further improvement of the
statistical errors, both for the form factor parameters and for the
measurement of $|V_{cb}|$.

\section{FORMALISM}
\label{sec:formalism}

\begin{figure}[ht]
\begin{center}
  \scalebox{0.8}{\includegraphics{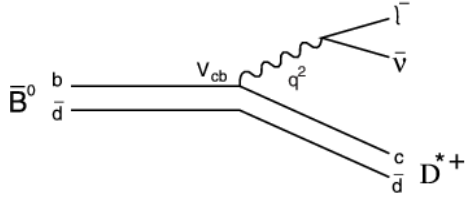}}
\end{center}
\caption{Quark-level diagram showing the weak interaction vertices in the decay \BztoDslnu.}
\label{fig:quarkleveldiag}
\end{figure}

This section outlines the formalism and describes the parameterization
used for the form factors. More details can be found in 
Refs.~\cite{ref:NeubertPhysReport,ref:Rich}.
The lowest order quark-level diagram for the decay \BztoDslnu is shown in Fig.~\ref{fig:quarkleveldiag}.

\subsection{Kinematic variables}

The decay \BztoDslnu is completely characterized by four variables,
three angles and $q^2$, the square of the momentum transfer from the
$B$ to the $D^*$ meson.

The momentum transfer is linearly related to another Lorentz-invariant
variable, called $w$, by

\begin{equation}
\label{eq:wdef}
w \equiv 
v_B \cdot v_{D^*} = { p_B \cdot p_{D^*} \over m_B m_{D^*}} =  {m_B^2 + m_{D^*}^2 -q^2 \over 2 m_B m_{D^*}},
\end{equation} 
where $m_B$ and $m_{D^*}$ are the masses of the $B$ and the $D^*$
mesons, $p_B$ and $p_{D^*}$ are their four-momenta, and $v_B$ and
$v_{D^*}$ are their four-velocities. In the $B$ rest frame the
expression for $w$ reduces to the Lorentz boost $\gamma_{D^*} =
E_{D^*}/M_{D^*} $.

The ranges of $w$ and $q^2$ are restricted by the kinematics of the
decay, with $q^2 \approx 0$ corresponding to
\begin{equation}
w_{max}=\frac{m_B^2+m_{D^*}^2}{2m_Bm_{D^*}}\approx1.504
\end{equation} 
and $w_{min}=1$ corresponding to
\begin{equation}
q_{max}^2=(m_B-m_{D^*})^2\approx10.69~(\gevcc)^2.
\end{equation}
 
\begin{figure}[ht]
\begin{center}
  \scalebox{0.8}{\includegraphics{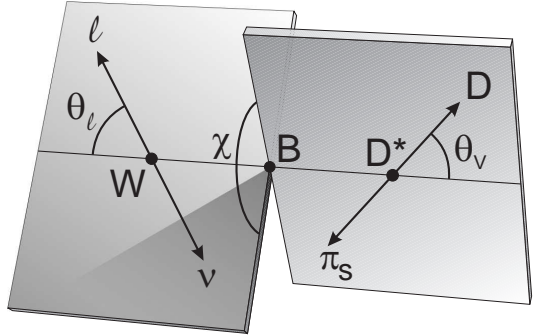}}
\end{center}
\caption{Kinematics of a \BztoDslnu  decay, mediated by an
  intermediate vector boson $W$; $\pi_s$ is the soft pion from the $D^*$ decay.  This diagram defines the three
  angles $\thetal$, $\thetav$ and $\angchi$.}
\label{fig:mesonleveldiag}
\end{figure}

In this analysis we only reconstruct the decay \dsptoDz,
where \DzKP, \DzKPPP or \DzKPPz.  The angular variables, shown in
Fig.~\ref{fig:mesonleveldiag}, are

\begin{itemize} 
  
\item{ $\thetal$, the angle between the direction of the lepton
    in the virtual $W$ rest frame and the direction of the virtual 
    $W$ in the $B$ rest frame, }
  
\item{ $\thetav$, the angle between the direction of the $D$ in the
    $D^*$ rest frame and the direction of the $D^*$ in the $B$ rest
    frame, }
  
\item{$\angchi$, the dihedral angle between the plane formed by the
    $D^{*}-D$ and the plane formed by the $W-\ell$ system. }

\end{itemize} 

\subsection{ Four-dimensional decay distribution}

The Lorentz structure of the \BztoDslnu decay amplitude can be
expressed in terms of three helicity amplitudes ($H_{+}$, $H_{-}$,
and $\H_{0}$), which correspond to the three polarization
states of the $D^*$ (two transverse and one longitudinal). 
For light leptons these amplitudes are expressed
\cite{ref:NeubertPhysReport,ref:Rich} in terms of the three form
factors $h_{A_1}$, $R_1$ and $R_2$:

\begin{equation}
H_i(w)=m_B\dfrac{R^*(1-r^2)(w+1)}{2\sqrt{1-2wr+r^2}}h_{A_1}(w)\tilde{H}_i(w),
\label{eq:HandHtilde}
\end{equation}
where
\begin{eqnarray}
\tilde{H}_{\mp} & = & \dfrac{\sqrt{1-2wr+r^2}\biggl(1\pm\sqrt{\dfrac{w-1}{w+1}
R_1(w)}\biggr)}{1-r} \\
\tilde{H}_0 & = & 1+\dfrac{(w-1)(1-R_2(w))}{1-r} 
\label{eq:HtildeandFF}
\end{eqnarray}
where $R^*=\dfrac{2\sqrt{m_B m_{\Dstar}}}{m_B+m_{\Dstar}}$ and
$r=\dfrac{m_{\Dstar}}{m_B}$.

The full differential decay rate in terms of the three helicity
amplitudes is~\cite{ref:NeubertPhysReport,ref:Rich}:

\begin{eqnarray}
\label{eq:totaldiffdecaywidth}
\dfrac{\mathrm{d\Gamma}(\BztoDslnu)}{\mathrm{d}w\mathrm{d\cos\theta_{\ell}}\mathrm{d\cos\theta_{V}}\mathrm{d\chi}} & = & \dfrac{6m_B m_{\Dstar}^2 r \sqrt{w^2-1}(1-2wr+r^2)}{8(4\pi)^{4}} \\ \nonumber
& \times & G_{F}^{2}|V_{cb}|^{2} \bigl[(1-\cos\theta_{\ell})^{2}\sin^{2}\theta_{V} H^2_{+}(w) \biggr. \\ \nonumber 
& + & (1+\cos\theta_{\ell})^{2}\sin^{2}\theta_{V} H^2_{-}(w)  \\ \nonumber
& + & 4\sin^{2}\theta_{\ell}\cos^{2}\theta_{V} H^2_{0}(w) \\ \nonumber
& - & 2\sin^2\theta_{\ell}\sin^2\theta_{V}\cos 2\chi H_{+}(w)H_{-}(w) \\ \nonumber    
& - &
4\sin\theta_{\ell}(1-\cos\theta_{\ell})\sin\theta_{V}\cos\theta_{V}\cos\chi
\times H_{+}(w)H_{0}(w) \\ \nonumber 
& + & \biggl. 4\sin\theta_{\ell}(1+\cos\theta_{\ell})\sin\theta_{V}\cos\theta_{V}\cos\chi \times H_{-}(w)H_{0}(w)\bigr],
\label{eq:totaldiffdeacaywidth}
\end{eqnarray}
where all three of the $H_i$ are functions of $w$. The
four-dimensional distribution of $w$, $\ctl$, $\ctv$, and $\chi$
described by Eq.~(\ref{eq:totaldiffdecaywidth}) is the physical observable from
which we extract the form factors. The normalization of this
distribution is directly related to $\Vcb$.

\subsection{Form-factors parameterizations}

The HQET does not predict the functional form of the form factors, and
one needs a parameterization of these form factors.  In the limit of
infinite mass for the $b$ and $c$ quark, $R_1$ and $R_2$ are simply
both equal to 1. Corrections to this approximation are calculated in
powers of $({\Lambda_{{\rm QCD}}}/{m_Q})$ and $\alpha_S$.

Parameterizations as a power expansion in $(w-1)$ have been proposed
in literature. The first linear parameterization is given by:
\begin{eqnarray}
  h_{A_{1}}(w) & = & h_{A_{1}}(1) (1-\rho^{2}(w-1)) \, .
\label{eq:linear}
\end{eqnarray}

A generic second order extension can be realized adding an extra
parameter $c$ as coefficient of the $(w-1)^2$ term:
\begin{eqnarray}
  h_{A_{1}}(w) & = & h_{A_{1}}(1) (1-\rho^{2}(w-1)+c(w-1)^2).
\label{eq:quadratic}
\end{eqnarray}

In both cases $R_1$ and $R_2$ are considered to be constants,
neglecting their $w$ dependence.

Among various predictions relating the coefficients of higher order
terms to that of the linear term, we decided to adopt the Caprini,
Lellouch and Neubert~\cite{ref:CLNpaper} one (in the following
referred to as CLN); the authors find the following functional forms
for the three HQET form factors:
\begin{eqnarray}
\label{eq:Cap}
h_{A_{1}}(w) & = &
h_{A_{1}}(1)\big[1-\rho^{2}_{h_{A_{1}}}z+(53\rho^{2}_{h_{A_{1}}}-15)z^{2}\big.
\\ \nonumber 
& & \big. -(231\rho^{2}_{h_{A_{1}}}-91)z^{3}\big], \\ \nonumber 
R_{1}(w) & = & R_{1}(1)-0.12(w-1)+0.05(w-1)^{2},  \\ \nonumber
R_{2}(w) & = & R_{2}(1)+0.11(w-1)-0.06(w-1)^{2},
\end{eqnarray}
where:
\begin{equation}
z=\dfrac{\sqrt{w+1}-\sqrt{2}}{\sqrt{w+1}+\sqrt{2}} \, .
\end{equation}

These functional forms are determined apart from three unknown
parameters, $\rho^{2}_{h_{A_{1}}}$, $R_{1}(1)$ and $R_{2}(1)$, which
must be extracted from the data.

It is important to notice that $h_{A_{1}}(1) \equiv \mathcal{F}(1)$,
corresponding to the value of the Isgur-Wise function evaluated at
zero recoil (using the notation commonly found in literature for
$d\Gamma/dw$). A recent lattice calculation~\cite{ref:auno} (including
a QED correction of 0.7\%) gives: $h_{A_{1}}(1) = \mathcal{F}(1) =
0.919^{+ 0.030}_{- 0.035}$.

\section{THE \babar\ DETECTOR}
\label{sec:babar}

The \babar\ detector is described in
detail in Ref.~\cite{ref:babar}.  The momenta of charged particles are
measured by a tracking system consisting of a five-layer silicon
vertex tracker (SVT) and a 40-layer drift chamber (DCH), operating in
a 1.5-T solenoidal magnetic field.  Charged particles of different
masses are distinguished by their energy loss in the tracking devices
and by a ring-imaging Cherenkov detector. Electromagnetic showers from
electrons and photons are measured in a CsI(Tl) calorimeter. Muons are
identified in a set of resistive plate chambers inserted in the iron
flux-return yoke of the magnet.

The analysis is based on a data sample of 79~\invfb\ recorded on the
\FourS\ resonance, and 9.6~\invfb\ recorded 40~MeV below it, with the
\babar\ detector \cite{ref:babar} at the PEP-II asymmetric-energy
\epem\ collider.  We use samples of GEANT Monte Carlo
(MC)~\cite{ref:geant4} simulated events that correspond to about three
times the data sample size. 

\section{RECONSTRUCTION AND EVENT SELECTION}
\label{sec:recoandsel}

The reconstruction of the events is the same as used in
Ref.~\cite{ref:bad776}, and the event selection is an improved version
of the one used in that paper.

We select events that contain a \dsp~candidate and an oppositely
charged electron or muon with momentum $1.2<p_{\ell}<2.4~\gevc$.
Unless explicitly stated otherwise, momenta are measured in the
\FourS\ rest frame, which does not coincide with the laboratory frame
due to the boost of the PEP-II beams.  In this momentum range, the
electron (muon) efficiency is about 90\% (60\%) and the hadron
misidentification rate is typically 0.2\% (2.0\%). We select \dsp\ 
candidates in the momentum range $0.5 < p_{D^*} < 2.5~\gevc$ in the
channel $\dsp \ra \Dz \psoft$, with the \Dz\ decaying to
$K^-\pi^+,~K^-\pi^+\pi^-\pi^+$, or $K^-\pi^+\pi^0$.  The charged
hadrons of the \Dz\ candidate are fitted to a common vertex and the
candidate is rejected if the fit probability is less than $0.1\%$. We
require the invariant mass of the hadrons to be compatible with the
\Dz\ mass within $\pm2.5$ times the experimental resolution. This
corresponds to a range of $\pm34\mevcc$ for the $\Dz \ra
K^-\pi^+\pi^0$ decay and $\pm17\mevcc$ for the other decays.
For the decay $\Dz \ra K^-\pi^+\pi^0$, we accept only candidates from
portions of the Dalitz plot where the square of the decay amplitude,
as determined by Ref.~\cite{ref:Dalitz}, is at least 10$\%$ of the
maximum it attains anywhere in the plot.  For the soft pion from the
\dsp\ decay, \psoft, the momentum in the laboratory frame must be less
than 450~\mevc , and the transverse momentum greater than 50~\mevc .
Finally, the lepton, the $\psoft$, and the \Dz are fitted to a common
vertex using a beam-spot constraint, and the probability for this fit
is required to exceed 1\%.

In semileptonic decays, the presence of an undetected neutrino
complicates the separation of the signal from background. We compute a
kinematic variable with considerable power to reject background by
determining, for each $B$-decay candidate, the cosine of the angle
between the momentum of the \Bzb\ and of the $\dsp\ell^-$ pair, under
the assumption that only a massless neutrino is missing:
\begin{equation*}
 \cos\TBY = \frac{2E_{\Bzb} E_{D^*\ell} - M^2_{\Bzb} - M^2_{D^*\ell} } { 2 p_{\Bzb} p_{\raisebox{-0.3ex}{\scriptsize $D^*\ell$}} }.
\end{equation*}
This quantity constrains the direction of the \Bzb\ to lie along a
cone whose axis is the direction of the $\dsp \ellm$ pair, but with an
undetermined azimuthal angle about the cone axis. The value of $w$
varies with this azimuthal angle; we take the average of the minimum
and maximum values as our estimator \omt for $w$.  This results in a
resolution of $0.04$ on \om .  We divide the sample into 10 bins in
\omt\ from 1.0 to 1.5, with the last bin extending to the kinematic
limit of 1.504.

The selected events are divided into six subsamples,
corresponding to the two leptons and the three \Dz\ decay modes.  In
addition to signal events, each subsample contains background events
from six different sources:
\begin{itemize}
\item combinatoric (events from $\BB$ and continuum in which at least
  one of the hadrons assigned to the \dsp\ does not originate from the
  \dsp\ decay);
\item continuum ($\dsp \ell^-$ combinations from $\epem \ra
  c\bar{c}$);
\item fake leptons (combined with a true \dsp);
\item uncorrelated background ($\ell$ and \dsp\ produced in the decay
  of two different $B$ mesons);
\item events from charged \B background, $\B^+ \rightarrow \Dstar \ellp
  \nul X$ (via $\B^+ \rightarrow \overline{D}^{**0} \ellp \nul$ or
  non-resonant
  $\B^+ \rightarrow \Dstarm \ellp \nul \pi^+$ production), and neutral \B
  background $\Bz\ra\Dstarm \ellp \nul X$, (via
  $\Bz\ra D^{**-} \ellp \nul$ or non-resonant
  $\Bz\ra\Dstarm \ellp \nul \piz$ production);
\item and correlated background events due to the processes $\Bzb \ra
  \dsp \bar{\nu} \tau^-,~\tau^- \ra \ellm X$ and $\Bzb \ra \dsp X_c,
  ~X_c \ra \ellm Y$.

\end{itemize}
We estimate the correlated background (which amounts to be less than
0.5\% of the selected candidates) from the Monte Carlo simulation
based on measured branching fractions \cite{ref:pdg04,ref:pdg05},
while we determine all others from data.  Except for the
combinatoric background, all other background sources exhibit a peak
in the $\dm = M_{\dsp} - M_{\Dz}$ distribution, where $M_{\dsp}$ and
$M_{\Dz}$ are the measured \dsp\ and \Dz\ candidate masses.

We determine the composition of the subsamples in each \omt\ bin in
two steps.  First we estimate the amount of combinatoric, continuum,
and fake-lepton background by fitting the \dm\ distributions in the
range $0.139<\dm<0.165~\gevcc$ simultaneously to three sets of events:
data recorded on resonance, data taken below the \FourS (thus
containing only continuum background), and data in which tracks that
fail very loose lepton-selection criteria are taken as surrogates for
fake leptons.
The distributions are fitted with the sum of three Gaussian functions
with a common mean and different widths to describe $\dsp \ra \Dz
\psoft$ decays and with empirical functions
\begin{equation}
  \label{eq:combshape}
  F_{comb}(\dm)= \frac{1}{N}\, \left[ 1- e^{\left(-\frac{\dm
      - \dm_0}{c_1}\right)}\right]\,
\left(\frac{\dm}{\dm_0}\right)^{c_2}
\end{equation}
based on simulation for the combinatoric background. The four
parameters of the Gaussian functions are common, while
the fraction of peaking background events and the parameters
describing the combinatoric background differ for the signal,
off-peak, and fake-lepton samples.

\begin{figure}
\begin{center}
  \scalebox{0.8}{\includegraphics{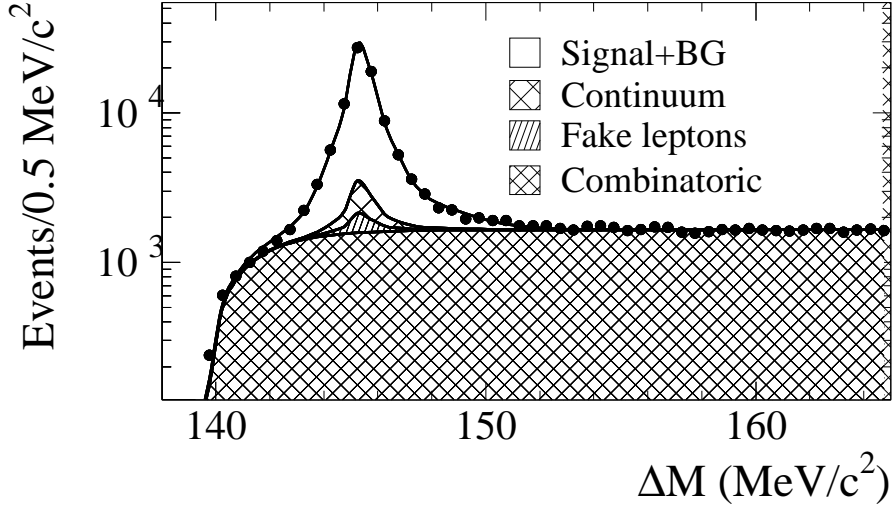}}
  \vspace*{-1.0cm}
\caption{Yields of on-resonance data (points) and the results of the fit (line)
  to the \dm\ distribution, with contributions from continuum,
  fake-lepton, and combinatoric-\dsp\ background events summed over
  all \omt\ bins. }
\label{fig:dm}
\end{center}
\end{figure}

Since the \dm\ resolution depends on whether or not the \psoft~track
is reconstructed only in the SVT or in the SVT and DCH, the fits are
performed separately for these two classes of events.  We rescale the
number of continuum and fake-lepton events in the mass range $0.143 <
\dm < 0.148~\gevcc$, based on the relative on- and off-resonance
luminosity and measured hadron misidentification probabilities.  The
fit is performed in multiple steps: the combinatoric background is
first fixed to the simulation prediction in order to optimize the fit
of the signal in the tail region around the peak. Then the peak shape
is fixed to the fitted one, and the combinatoric background is left
free to float, to allow for possible small deviations from the
simulation.

In the subsequent analysis we fix the fraction of combinatoric,
fake-lepton, and continuum events in each \omt\ bin to the values
obtained following the procedure described above. Figure~\ref{fig:dm}
shows the \dm\ fit results for the on-resonance data.

In a second step, we fit the $\cos\TBY$ distributions in the range
$-10 < \cos\TBY < 5$ and determine the signal contribution and the
normalization of the uncorrelated and $B \rightarrow \Dstar \ellp
  \nul X$ background events.
Neglecting resolution effects, signal events meet the constraint
$|\cos\TBY|\leq1$, while the distribution of $B \rightarrow
\Dstar \ellp \nul X$ events
extends below $-1$, and the distribution of the uncorrelated background events
is spread over the entire range considered.

\begin{figure}
\begin{center}
  \scalebox{0.8}{\includegraphics{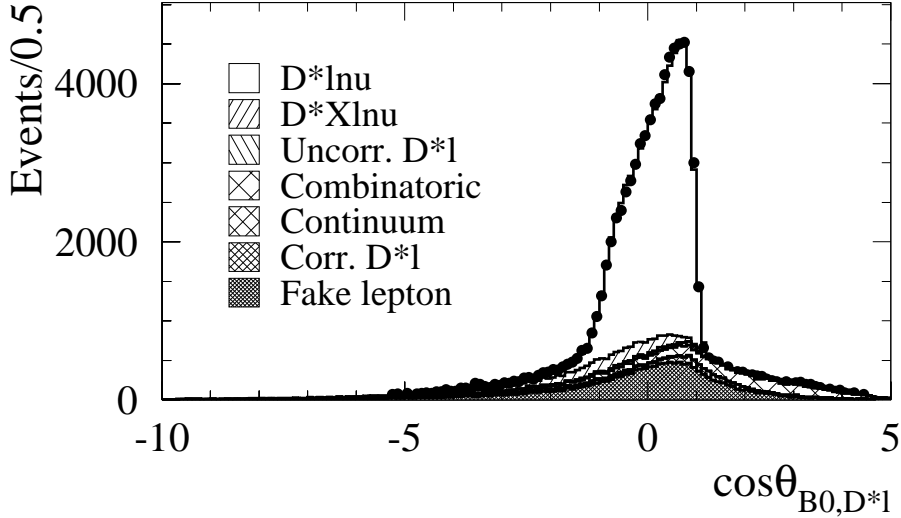}}
  \vspace*{-1.0cm}
\caption{ Yields of on-resonance data (points) and the results of the fit (histograms) to the $\cos\TBY$  
  distribution, summed over all \omt\ bins.}
\label{fig:ctBY}
\end{center}
\end{figure}

We perform the fit separately for each \omt\ bin, with the individual
shapes for the signal and for each of the six background sources taken
from MC simulation, specific for each of the six subsamples.  Signal
events are generated with the CLN form factor parameterization, tuned
to the results from CLEO~\cite{ref:CLEOff}. Radiative decays (\BtoDs
$\gamma$) are modeled by PHOTOS \cite{ref:photos} and are treated as
signal. $\B \to D^{**} \ell \nu$ decays involving orbitally excited
charm mesons are generated according to the ISGW2
model~\cite{ref:IGSW}, and decays with non-resonant charm states are
generated following the prescription in Ref.~\cite{ref:Goity}.  To
reduce the sensitivity to statistical fluctuations we require that the
ratio of $B \rightarrow \Dstar \ellp \nul X$ and of uncorrelated background events to the signal
be the same for all three \Dz\ decay modes and for the electron and
muon samples.  Fit results are shown in Fig.~\ref{fig:ctBY}.  In
total, there are 68,840 events in the range $|\cos\TBY|<1.2$.  The
average fraction of signal events is
$(76.7\pm0.3)$\%, where the error is only statistical.

\section{THE EXTRACTION OF \boldmath{\Vcb} AND THE FORM FACTOR PARAMETERS}
\label{sec:analysis}

The approach followed in this analysis to simultaneously measure \Vcb
and the three form factor parameters is to extend the one-dimensional
fit procedure of Ref.~\cite{ref:bad776} to several one-dimensional
distributions, while fully accounting for the mutual
correlations, using the same sample of selected events. This permits 
disentangling the effects coming from each of the parameters, and to
measure all of them.

In principle, the multi-dimensional phase space could be divided into
several independent bins, and the fraction of signal events in each of
them could be determined as described in the previous section. In
practice, the need for a detailed bin segmentation would imply serious
statistical limitations in most of the bins. For this reason, we
instead consider projections of the data into the most significant
variables ( $w$, \ctl, \ctv). We divide each projection into ten bins,
and determine the amount of signal events in each bin applying the
procedure described above. We then fit the resulting projections to
extract the form factors and \Vcb. We account for the correlations
between bins of different projections by noting that the covariance
between the content of two bins in different one-dimensional
distributions is given by the common number of events in these bins,
while it is zero for bins in the same distribution.

\begin{figure*}[htp]
\begin{center}
\begin{tabular}{ccc}
\includegraphics[width=0.25\textwidth,clip=]{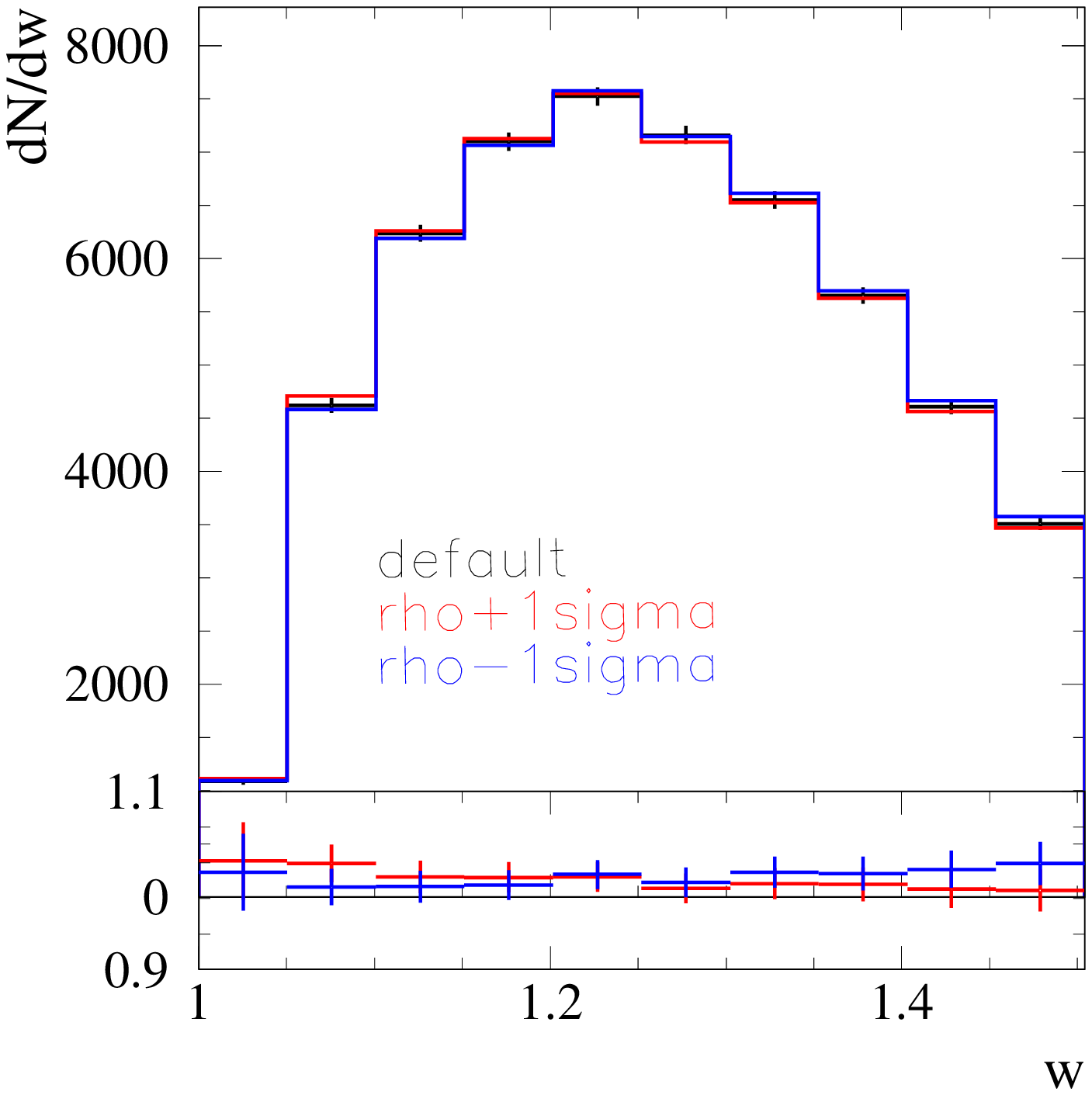} &
\includegraphics[width=0.25\textwidth,clip=]{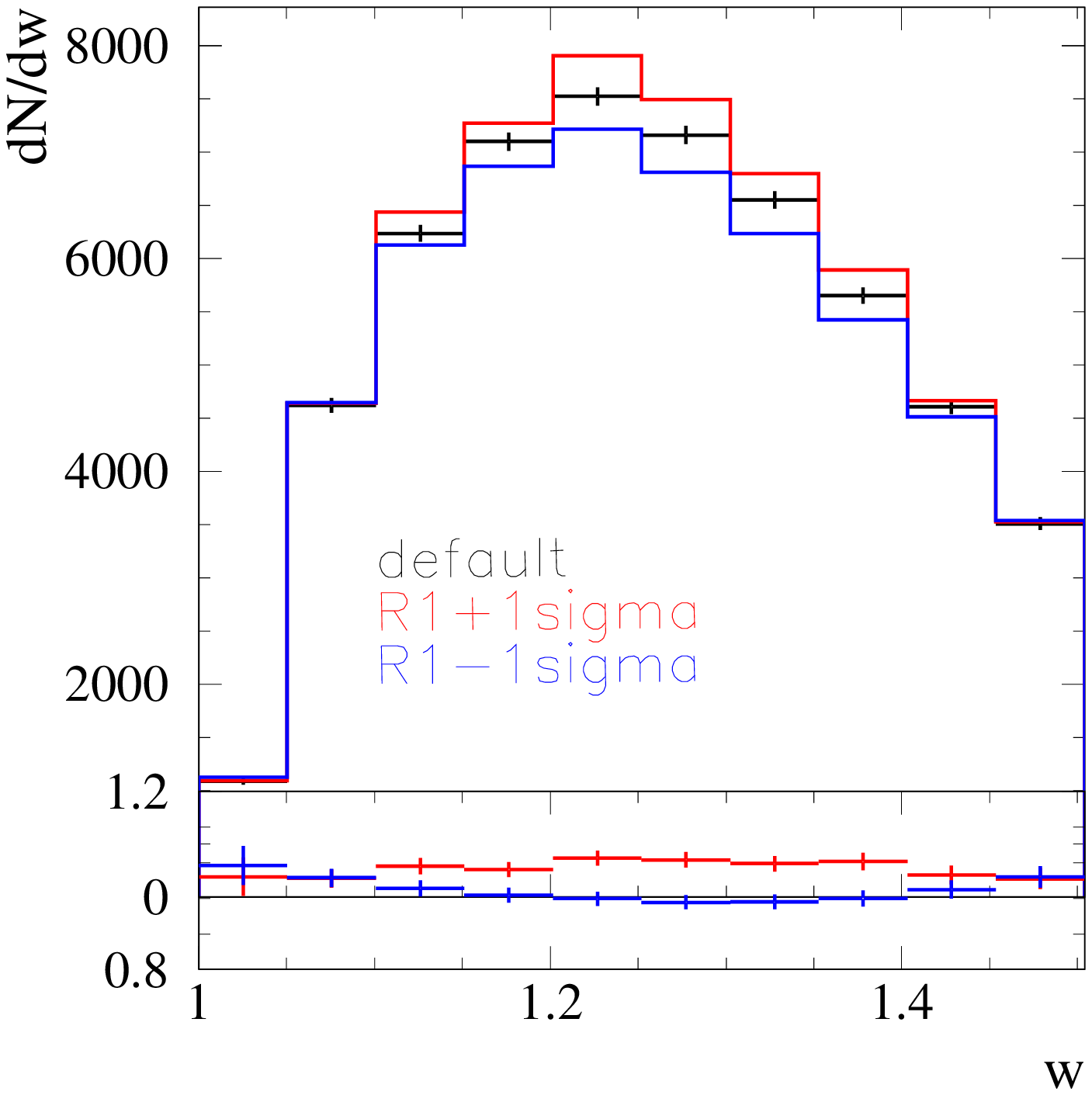} & 
\includegraphics[width=0.25\textwidth,clip=]{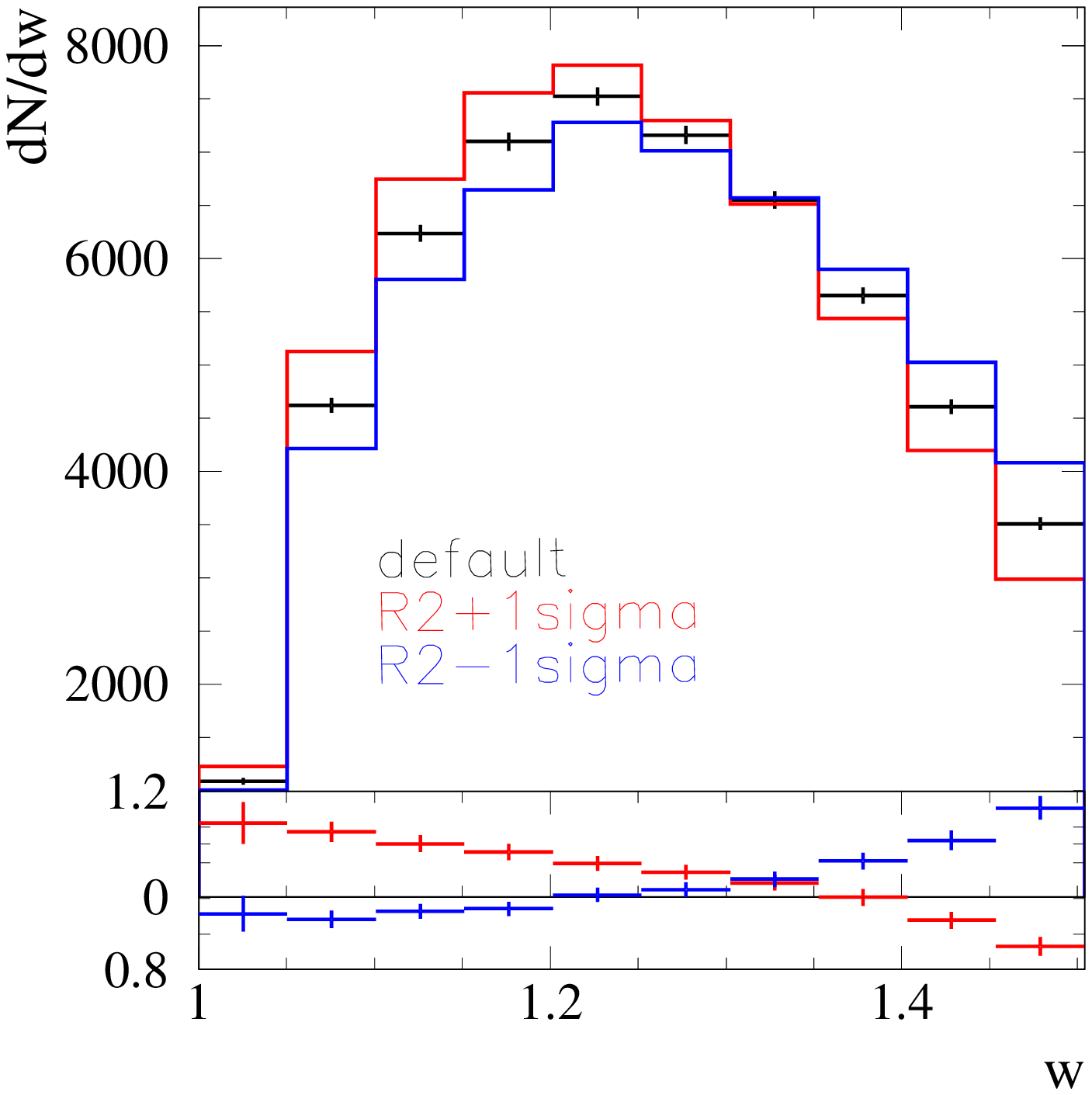} \\
\includegraphics[width=0.25\textwidth,clip=]{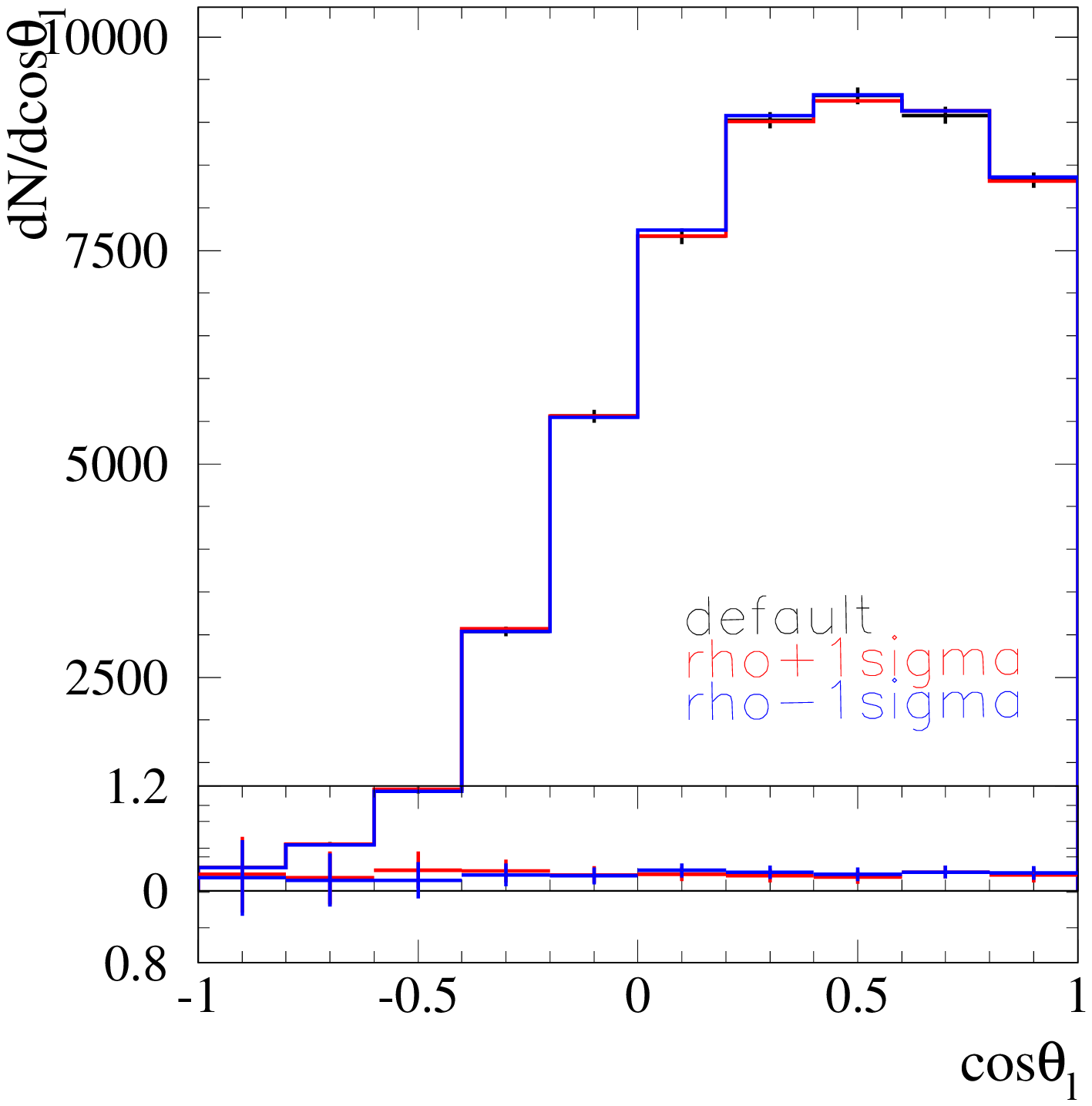} &
\includegraphics[width=0.25\textwidth,clip=]{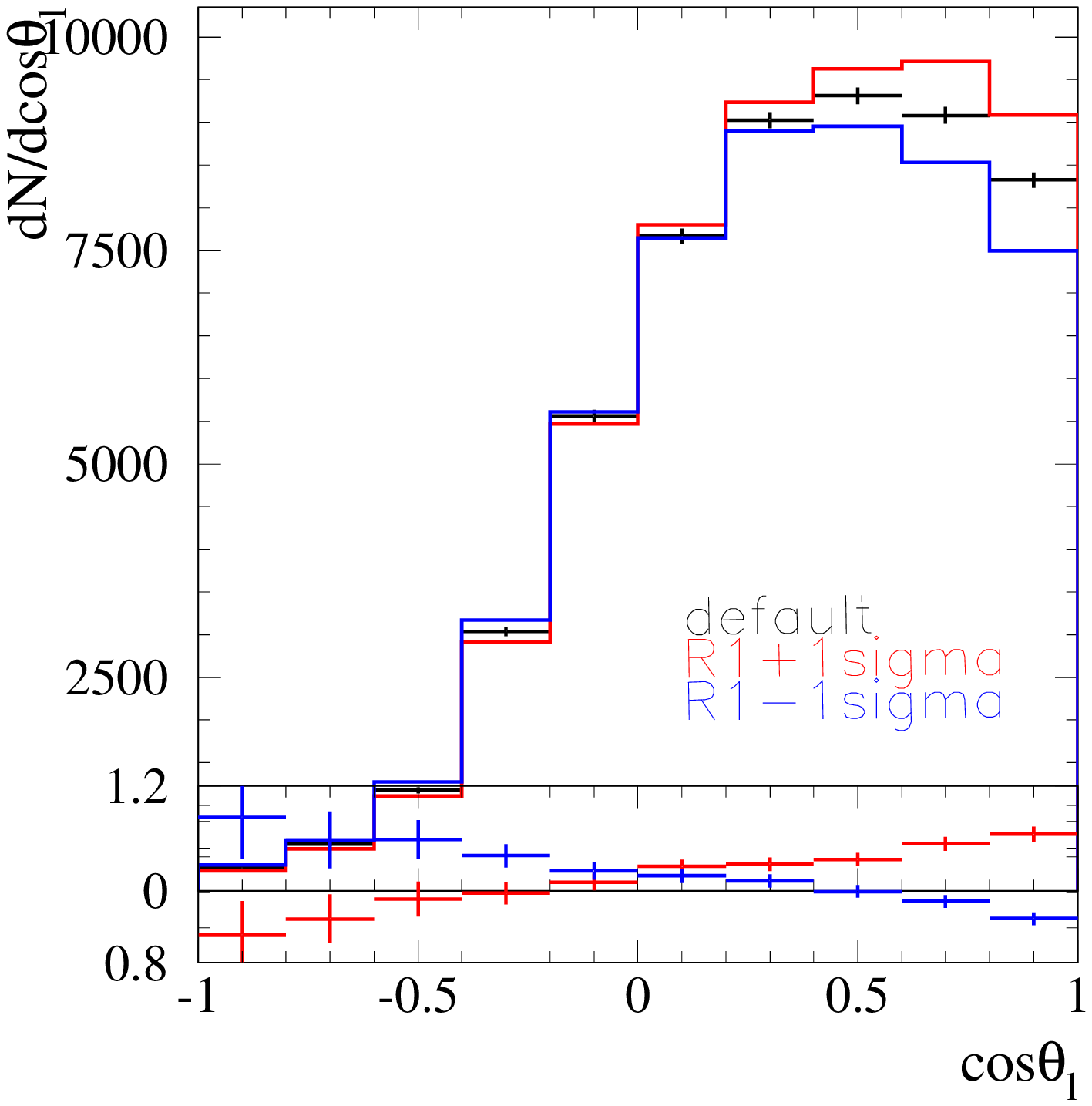} & 
\includegraphics[width=0.25\textwidth,clip=]{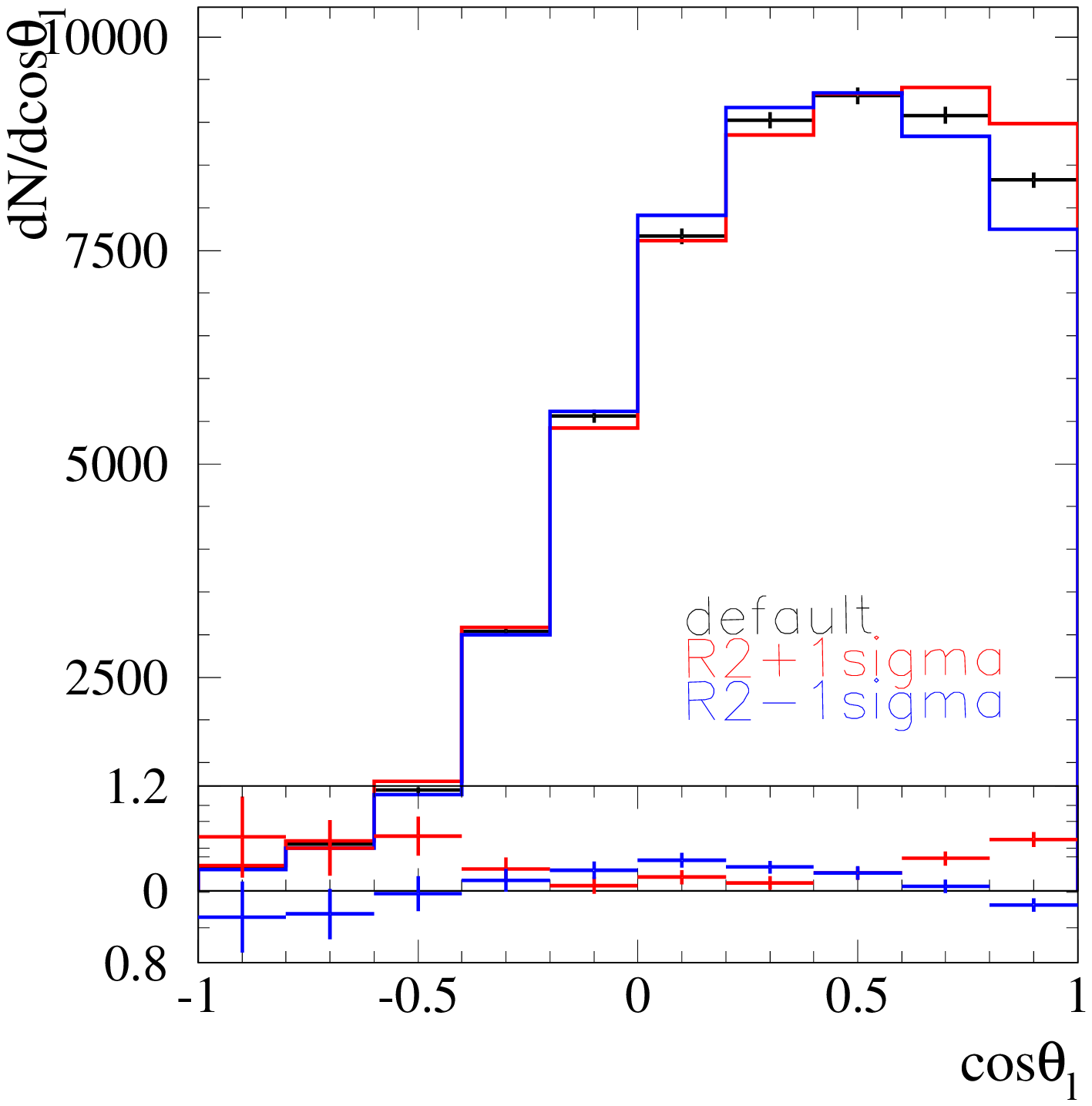} \\
\includegraphics[width=0.25\textwidth,clip=]{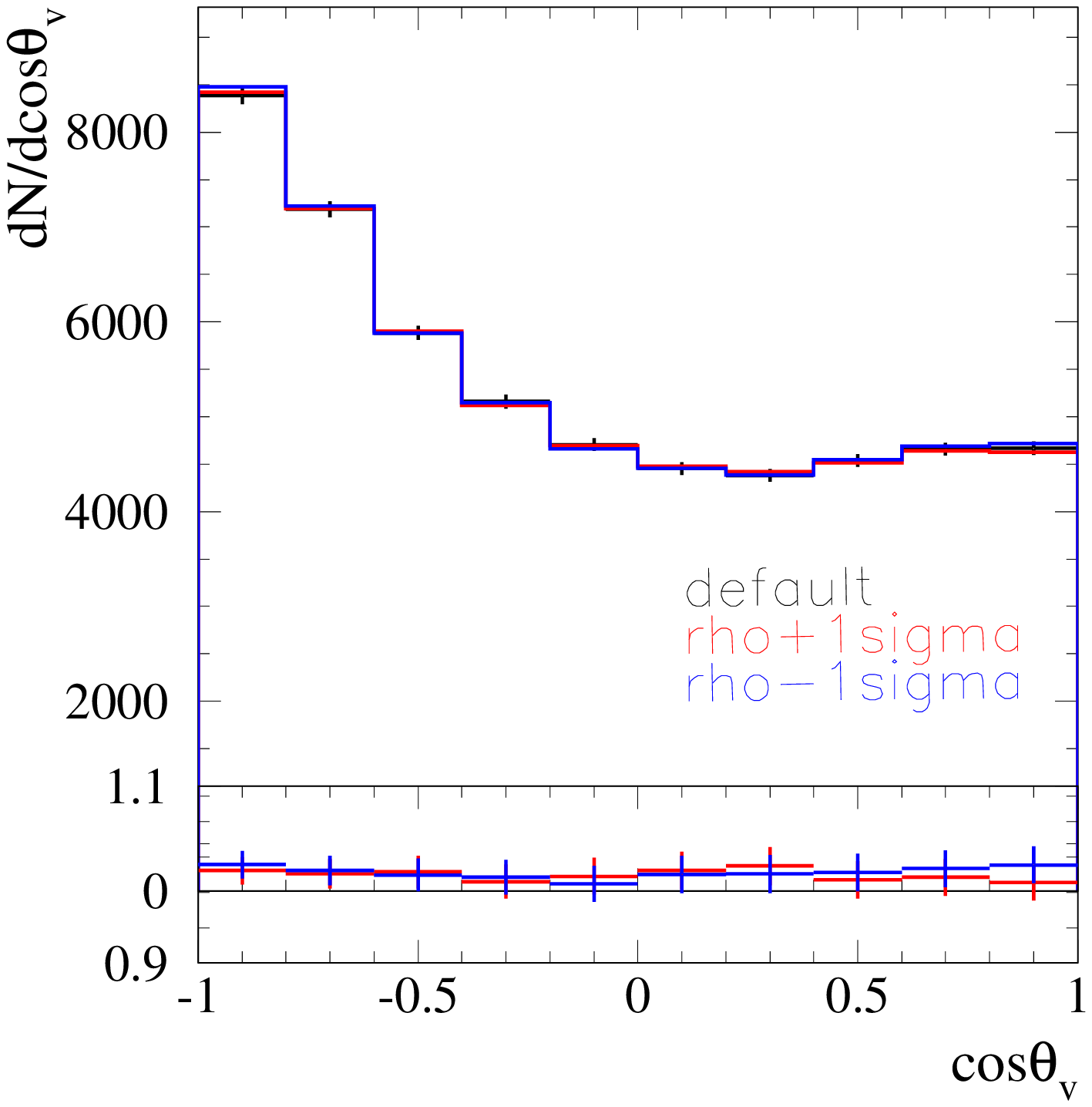} &
\includegraphics[width=0.25\textwidth,clip=]{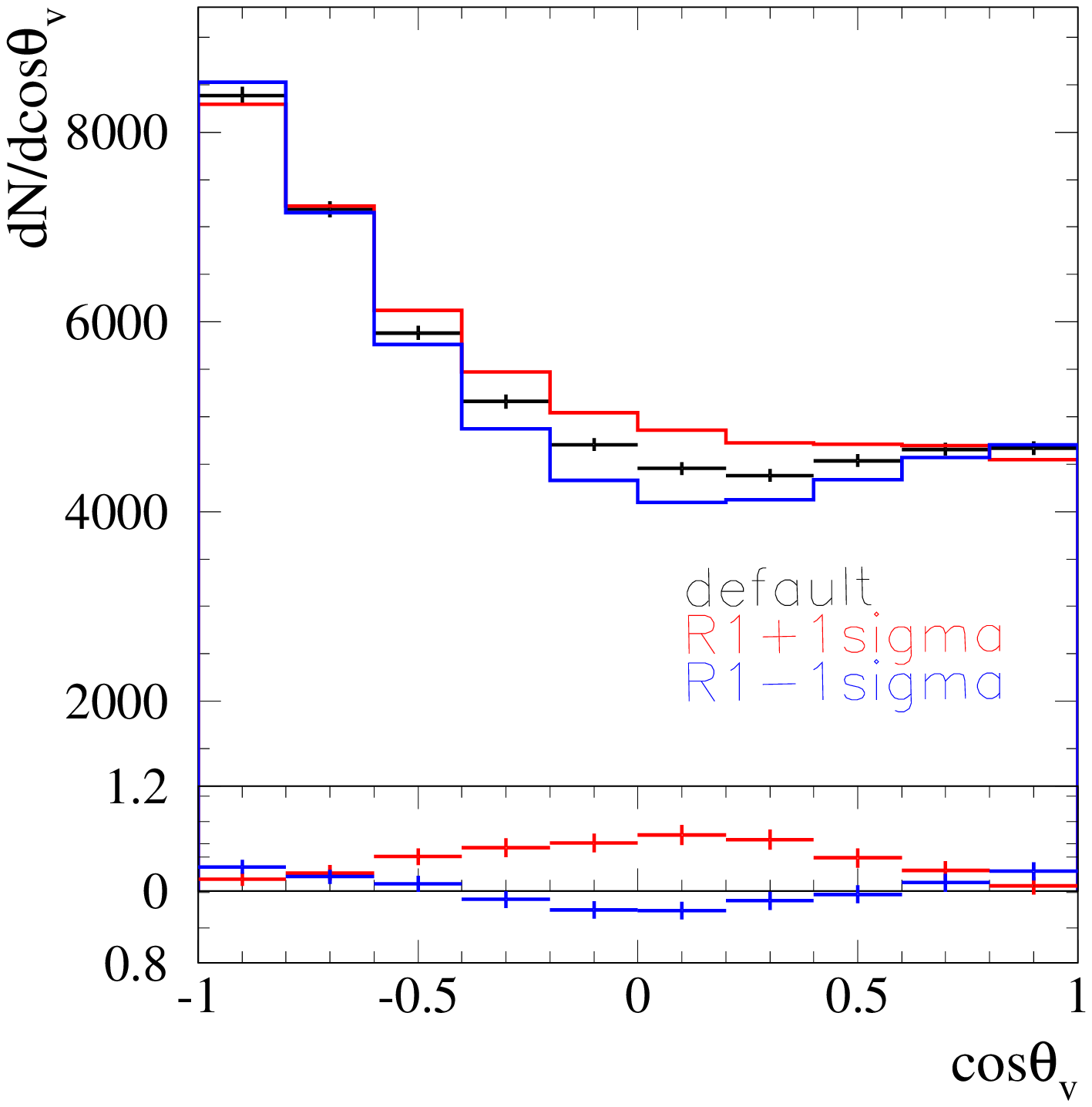} & 
\includegraphics[width=0.25\textwidth,clip=]{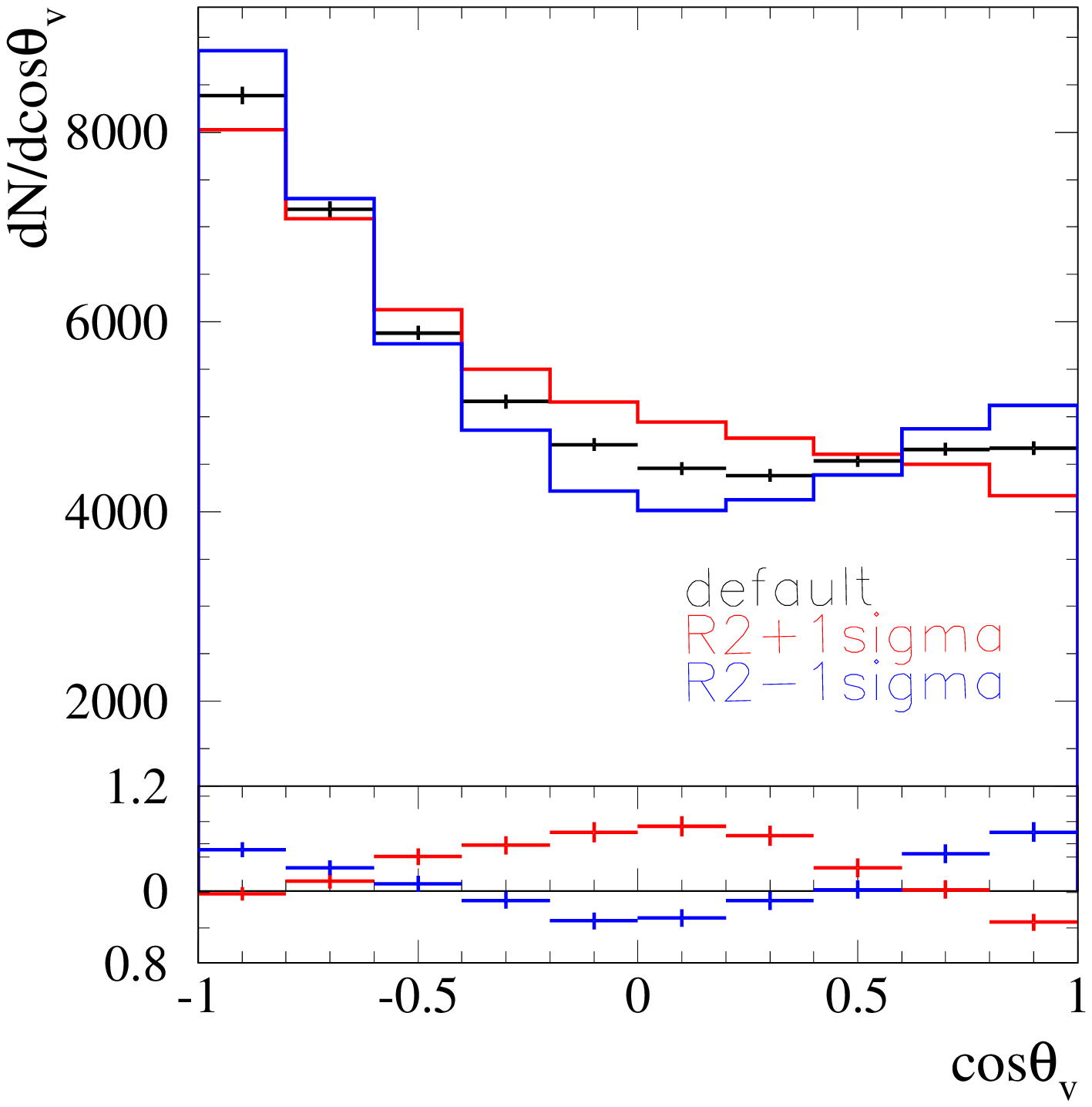} \\
\includegraphics[width=0.25\textwidth,clip=]{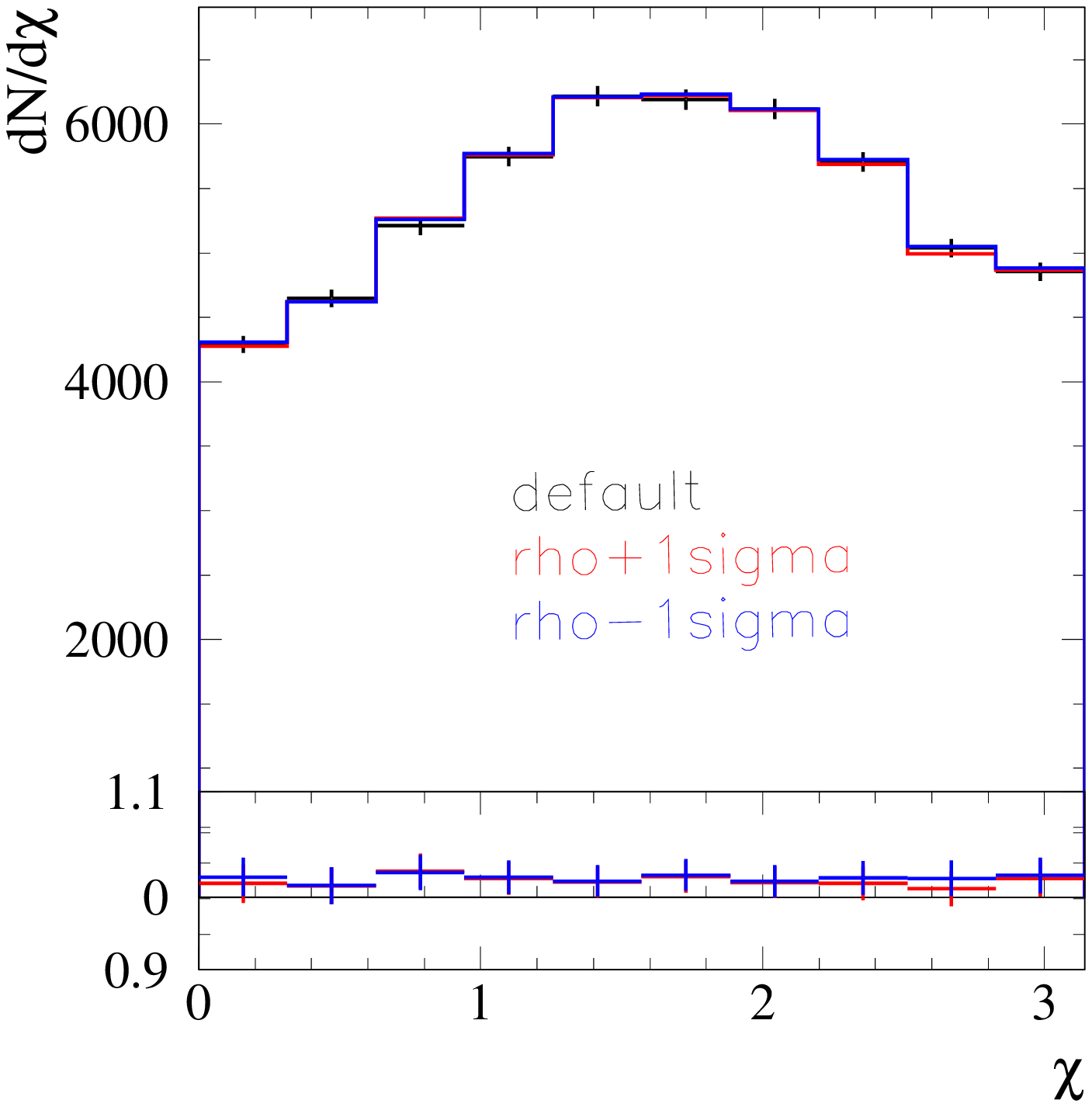} &
\includegraphics[width=0.25\textwidth,clip=]{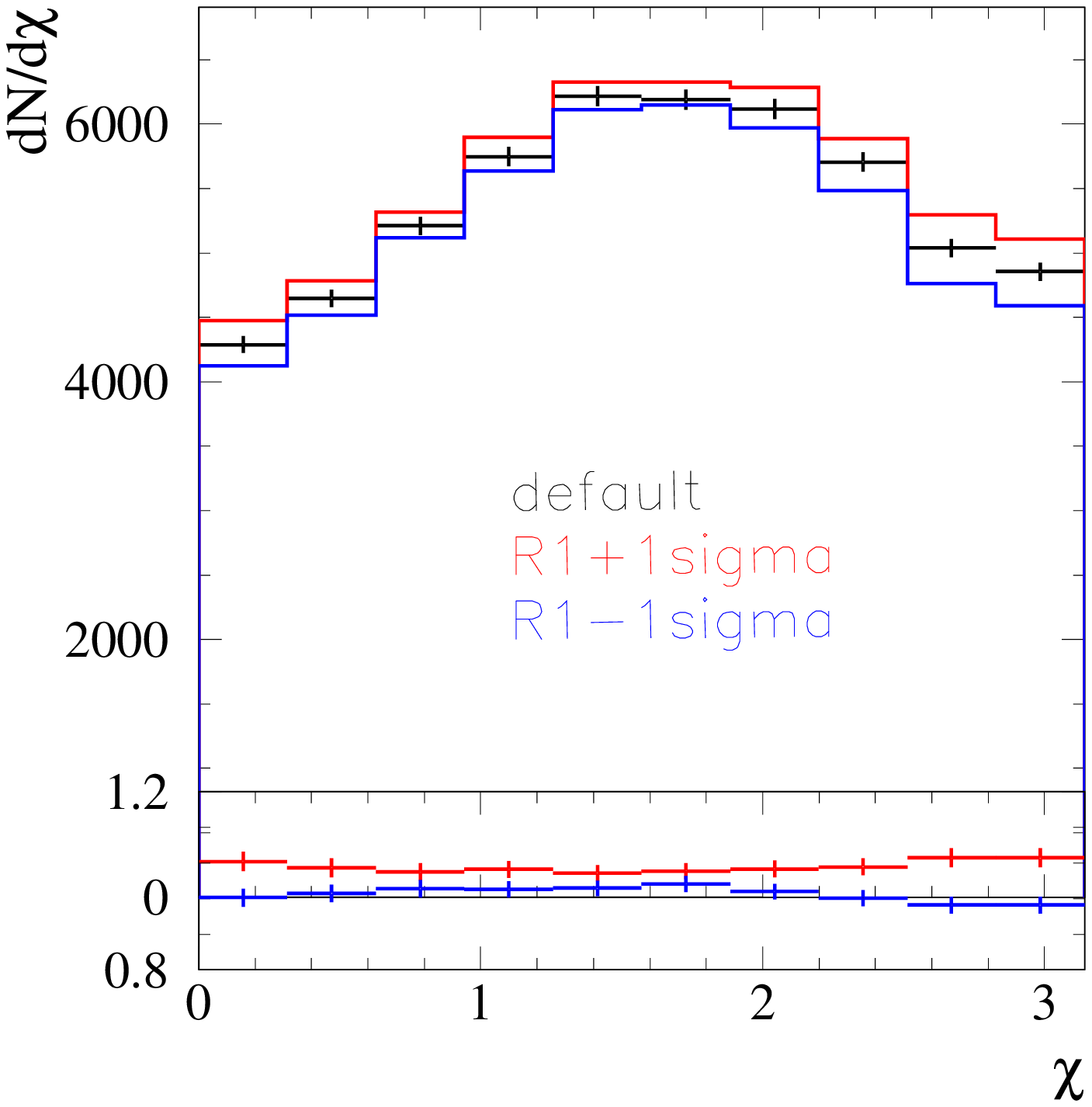} & 
\includegraphics[width=0.25\textwidth,clip=]{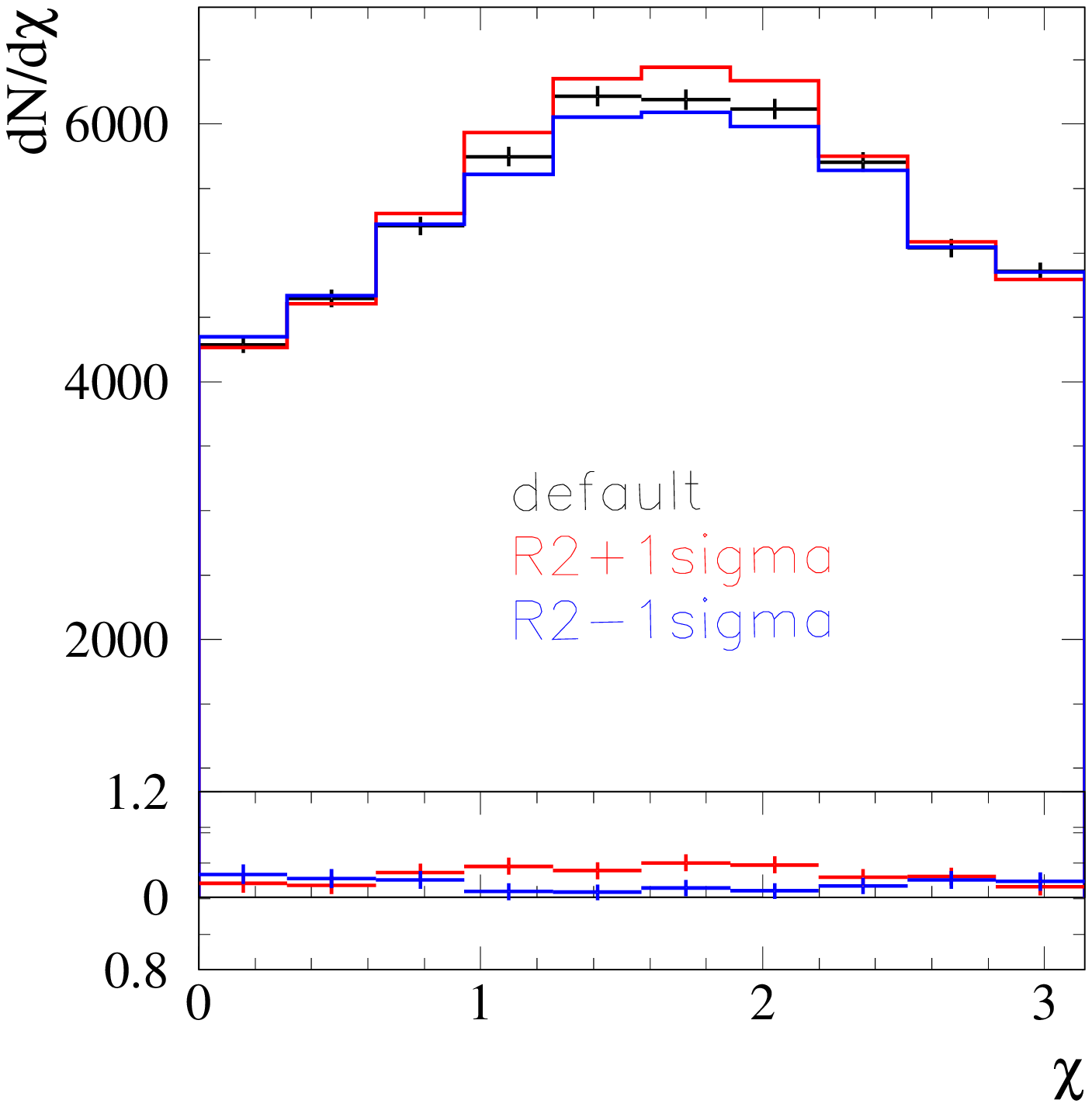} 
\end{tabular}
\caption{ Sensitivity of the $w$, $\ctl$, $\ctv$ and $\chi$ variables
  to the variations of the three form factor parameters $\rho^2$,
  $R_1$ and $R_2$. The variations considered are 1 $\sigma$ for
  $\rho^2$, $R_1$, and $R_2$ using the measured value from~\cite{ref:bad776}
  for $\rho^2$ and from Ref.~\cite{ref:CLEOff} for the others.}
\label{fig:sensitivity}
\end{center}
\end{figure*}

Any kinematic observable sensitive to the form factor parameters can
be used, the sensitivity to \Vcb being already present in the $w$
distribution alone. We have studied the distributions of the kinematic
observables describing the decay, as discussed in
Sec.~\ref{sec:formalism}. We have verified that the distribution of
$\chi$ is practically insensitive to the form factor parameter
changes, therefore we have chosen  $w$, $\ctl$ and $\ctv$ as set of
observables to be included in the fit. Figure~\ref{fig:sensitivity} shows
the sensitivity of each of the 4 kinematical variables describing the
\BztoDslnu decay to the form factor parameters.

The value of \Vcb and the form factor parameters are extracted from a
least squares fit to the set of measurements given by the measured
contents of bins of $w$, $\ctl$ and $\ctv$, according to the selection
described in Sec.~\ref{sec:recoandsel} and the cut
$|\cos\TBY| < 1.2$. The $w$ and $\ctv$ distributions are
divided into ten equal size bins. The $\ctl$ distribution is divided in
the analysis in 10 non-equal size bins, whose boundaries are $[-1.0,
-0.4, -0.2, 0.0, 0.2, 0.4, 0.5, 0.6, 0.7, 0.8, 1.0]$, chosen according
to the strong variations in the shape, and therefore to the
available statistics.

\subsection{The fit method}

The $\chi^2$ function to be minimized is an extension of the
one used in Ref.~\cite{ref:bad776}. Defining $n$ as the total number
of observable bins, $N_i^{data}$ the total
number of observed events in bin $i$, $N_i^{bkg}$ the number of
estimated background events for that bin, and $N_i^{MC}$ the number of
simulated signal events falling in that bin, the $\chi^2$ can be
written as:

\begin{eqnarray}
  \chi^2 & = & \sum_{i = 1}^{n} \sum_{j = 1}^{n} \biggl (N_i^{data} - N_i^{bkg} - \sum_{k=1}^{N_i^{MC}} \mbox{W}_i^k
  \biggr ) \times \\ \nonumber 
  & & C_{ij}^{-1} \biggl ( N_j^{data} - N_j^{bkg} - \sum_{k=1}^{N_i^{MC}} \mbox{W}_j^k \biggr ) 
\label{eq:chi2}
\end{eqnarray}
where $\mbox{W}_i^k$ is the weight assigned to the $k$-th simulated signal event
falling into bin $i$ in order to evaluate the expected signal yield
as a function of the searched parameters, and $C_{ij}$ is the
covariance matrix element for the bin pair $ij$.

Each weight $\mbox{W}_i^k$ is the product of four weights,
$\mbox{W}_i^k = \mbox{W}^{\mathcal{L}} \mbox{W}_i^{\epsilon,k}
\mbox{W}_i^{S,k} \mbox{W}_i^{ff,k}$. 
The factor $\mbox{W}^{\mathcal{L}}$ accounts
for the normalization of data and simulation samples. It depends on
the total number of $B\bar{B}$ events, $N_{B\bar{B}} = (85.9 \pm 0.9)
\times 10^6$, on the fraction of $B^0 \bar{B}^0$ events, $f_{00} =
0.489 \pm 0.012$~\cite{ref:pdg04}, on the branching fraction
$\mathcal{B}(D^{*+} \rightarrow D^0 \pi^+) = 0.677 \pm
0.005$~\cite{ref:pdg04}, and on the $B^0$ lifetime $\tau_{B^0} = 1.532
\pm 0.009$ ps~\cite{ref:pdg04}. 
The factor $\mbox{W}_i^{\epsilon,k}$ accounts for
differences in reconstruction and particle-identification efficiencies
predicted by the Monte Carlo simulation and measured with data, as a
function of particle momentum and polar angle. 

The factor
$\mbox{W}_i^{S,k}$ accounts for potential small residual differences
in efficiencies for the six data subsamples (3 $D^0$ decay modes and 2
leptons) and allows for adjustment of the $D^0$ branching fractions,
properly dealing with the correlated systematic uncertainty. It is the
product of several scale factors that are floating parameters in the
fit, each constrained to an expected value with a corresponding
experimental error. To account for the uncertainty in the
multiplicity-dependent tracking efficiency, we introduce a factor
$\mbox{W}_{trk}^{S} = 1 + N_{trk} (1 - \delta_{trk})$, where $N_{trk}$
is the number of charged tracks in the \Dsl candidates in each sample
and $\delta_{trk}$ is constrained to one within the estimated
uncertainty in the single-track efficiency of $0.8$\%. Similarly,
multiplicative correction factors are introduced to adjust lepton
($\delta_{PID \, e}$, $\delta_{PID \, \mu}$) efficiencies, kaon
efficiencies in the
different selections used for different \Dz decay channels ($\delta_{PID \, K
  \, tight}$, $\delta_{PID \, K \, not \, a \, pion}$), and $\pi^0$
($\delta_{\pi^0}$) efficiencies, within the total estimated uncertainties of these
efficiencies, and $D^0$
branching fractions ($\delta_{BR \, D^0 \, K\pi}$, $\delta_{BR \, D^0
  \, K\pi\pi\pi}$, $\delta_{BR \, D^0 \, K\pi\pi^0}$). 
Correlations between the branching fractions are taken into account in
the constraint through the covariance matrix $C_{BR}$ (given by the
uncertainties and the correlation matrix obtained
from Ref.~\cite{ref:pdg04,ref:pdg05}). Therefore $\mbox{W}_i^{S,k}$
can be written as:
\begin{equation} 
\mbox{W}_i^{S,k} = \delta_{PID \, e}\delta_{PID \,
  \mu}\delta_{PID \, K}^{i,k}(1 +
  N_{trk}^{i,k}(1-\delta_{trk}))\delta_{\pi^0}^{i,k}\delta_{BR \, D^0}^{i,k}
\label{eq:weight}
\end{equation}
where the kaon and $\pi^0$ efficiency corrections, the decay
multiplicity $N_{trk}$ and the $D^0$ branching ratio correction depend
on the particular event $k$ of the bin $i$ considered (because of the
different decay multiplicity, branching ratio, kaon selector and
presence of $\pi^0$).  The addition of the
constraints effectively adds extra terms to the above $\chi^2$
expression for each of the corrective factors
$\delta$. The complete $\chi^2$ function used in the fit therefore is:
\begin{eqnarray}
  \chi^2 & = & \sum_{i = 1}^{n} \sum_{j = 1}^{n} \biggl (N_i^{data} -
  N_i^{bkg} - \sum_{k=1}^{N_i^{MC}} \mbox{W}_i^k \biggr ) \times 
  C_{ij}^{-1} \biggl ( N_j^{data} - N_j^{bkg} -
  \sum_{k=1}^{N_i^{MC}} \mbox{W}_j^k \biggr ) \\ \nonumber
  & & + \frac{(1 - \delta_{PID \, e})^2}{\sigma_{PID \, e}^2} +
  \frac{(1 - \delta_{PID \, \mu})^2}{\sigma_{PID \, \mu}^2} \\ \nonumber
  & & + \frac{(1
    - \delta_{PID \, K \, tight})^2}{\sigma_{PID \, K \, tight}^2} +
  \frac{(1 - \delta_{PID \, K \, not \, a \, pion})^2}{\sigma_{PID K
      \, not \, a \, pion}^2} \\ \nonumber
  & & + \frac{(1 - \delta_{trk})^2}{\sigma_{trk}^2} + \frac{(1 -
    \delta_{\pi^0})^2}{\sigma_{\pi^0}^2} \\ \nonumber 
  & & + \sum_{m = K\pi, K\pi\pi\pi,
    K\pi\pi^0} \sum_{n = K\pi, K\pi\pi\pi, K\pi\pi^0} \delta_{BR \,
    D^0 \, m} \times 
    C_{BR \, D^0 \, mn}^{-1} \delta_{BR \, D^0 \, n}
\label{eq:chi2full}
\end{eqnarray}
The addition of these extra terms allows us to fit all
subsamples simultaneously fully taking into account the correlated systematic
uncertainties, and effectively propagates these uncertainties through
the weights to the uncertainty on the free parameters of the fit.

The fourth factor $\mbox{W}_i^{ff,k}$ accounts for the dependence of
the signal yield on the parameters to be fitted. The \Vcb dependence
is trivially given by the ratio $|V_{cb}|^2/|V_{cb}^{MC}|^2$, where
the denominator is the actual value used in the simulation, derived
from the branching ratio used for the decay \BztoDslnu.  In an
analogous way the dependence on the form factor parameters ($\rho^2$, $R_1$, and $R_2$)
is given by the ratio of the differential decay rate of
Eq.~(\ref{eq:totaldiffdecaywidth}) evaluated at the variable values
(in the chosen fitting model) and at the simulation values (where the linear
model was used).

One important point is that not all bins of all used
observables can be used at the same time: in fact, once all bins
belonging to one observable are used, the overall normalization is
fixed, therefore for the other observables one of the bins has to be
dropped, since its content and covariance matrix part can be written
as linear combination of all others. The choice of the bins to be
dropped is arbitrary, and it has
been verified that by dropping different bins exactly the
same numerical results were obtained (within the numerical precision
of the fit).

Summarizing, three observables are used in the fit, $w$,
$\cos\theta_{\ell}$, and $\cos\theta_{V}$, whose ranges are divided into
ten bins. One bin is excluded from the fit for the last two
observables, for a total of 28 bins used in the $\chi^2$.

This fit procedure has been tested on a signal toy Monte Carlo, and no
evidence for systematic biases has been observed.

\subsection{The observables' covariance matrix}

A direct consequence of this method is that the covariance matrix for
the measurements corresponding to the bins used is not diagonal, since
the bins are not all statistically independent.
The total covariance matrix is the sum of 3 separate matrices: one for
the measured data yields, one for the estimated signal yield, and one
for the evaluated background.
The diagonal elements of the matrices are given by the uncertainty on
the bin content itself: the covariance of bins belonging to the same
observable is zero, the covariance of bins from different observables
is the variance of the number of events that is common between the bins.

The observed data matrix is built assuming the bin contents and their
intersections obeying a Poisson distribution, therefore the variance
of a bin content or of the content of the intersection of two bins is
the content itself. The estimated signal matrix is built in an
analogous way, where the variance of a number of weighted events $n$
is approximated by the sum of the squared weights $\sum_{i = 1,n}
w_i^2$.

The calculation of the background covariance matrix is less
straightforward. The diagonal part is simply the estimated variance of
the measured background, according to the whole procedure described in
Sec.~\ref{sec:recoandsel}.  The background extraction
procedure does not directly determine the
common number of events between two bins, because this procedure is
based on a complex sequence of shapes and yield fits.

The solution adopted is to use the number of common
background events between two bins as predicted by the simulation,
after it has been corrected for tracking and PID
efficiencies, and rescaled in such a way as to get for each bin of one of 
the studied kinematic observables (one among $w, \ctl, \ctv, \chi$)
the background amount estimated from the data-based background evaluation. 
In this estimation procedure also the $\chi$ projection is used, in
order to provide an additional check on the background evaluated from
the three distributions used in the fit. 

Since the total integral of the background measured is not exactly the
same when evaluated for different kinematic observables, it has to be
fixed to a given value. The average normalization 
determined from the four kinematic observables describing the event is used. 

The spread of the background normalization values is found to be about
two times larger than the estimated uncertainty from the background
procedure error propagation. This is due to the fact that the
uncertainties on the signal and background shapes are not accounted
for in the background estimate.
Taking into account the structure of the distribution of
the four integrals, split into two subgroups of compatible
values, the global background uncertainty is estimated to be the
average of the maximum and minimum difference between these two groups
({\it i.e.}, $w$ and \ctl on one side, \ctv and $\chi$ on the other). The
background uncertainties are suitably rescaled in order to give the
uncertainty on the integral equal to the one estimated as explained.

\section{THE FIT RESULTS AND SYSTEMATIC UNCERTAINTIES}
\label{sec:results}

\subsection{Results of the fit}

The analysis result is obtained using the background  
covariance matrix evaluated from the measurement in the variable
$w$. Table~\ref{tab:stdresu} shows the best fit 
result. The uncertainties have been determined by the {\tt
  MINOS}~\cite{ref:minuit}
algorithm, whose output has been symmetrized. The 
$\delta$ coefficients describing systematic uncertainties not common to
all events are found to be consistent with 1 within the estimated 
uncertainties. The contribution to the overall uncertainty
due to statistical fluctuations of the data and simulated samples has been
obtained by recomputing the covariance matrix of the fit after fixing all
the $\delta$ parameters at their best fit values.

\begin{figure}[htp]
\begin{center}
\begin{tabular}{cc}
\includegraphics[width=0.50\textwidth,clip=]{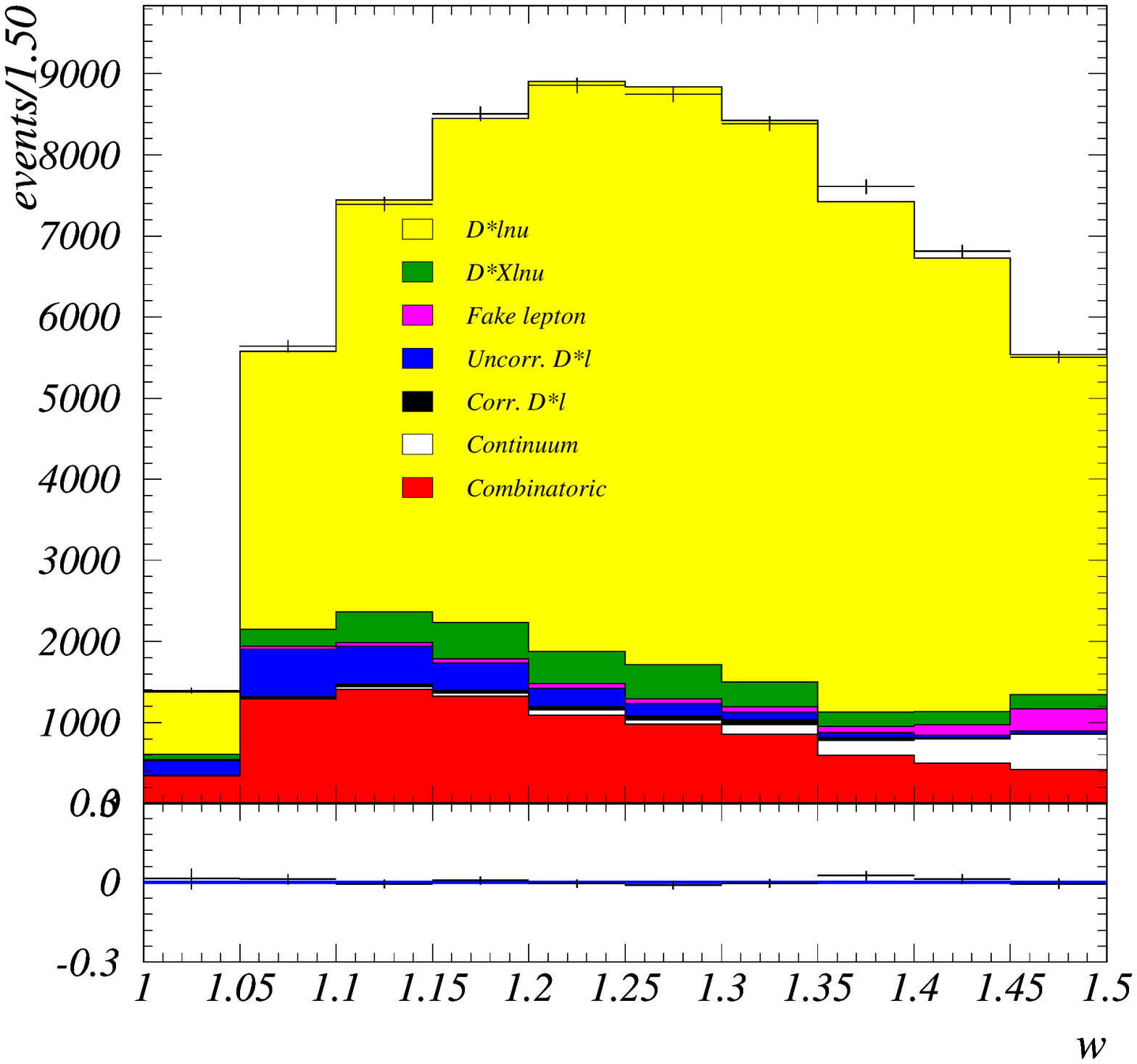}
&
\includegraphics[width=0.50\textwidth,clip=]{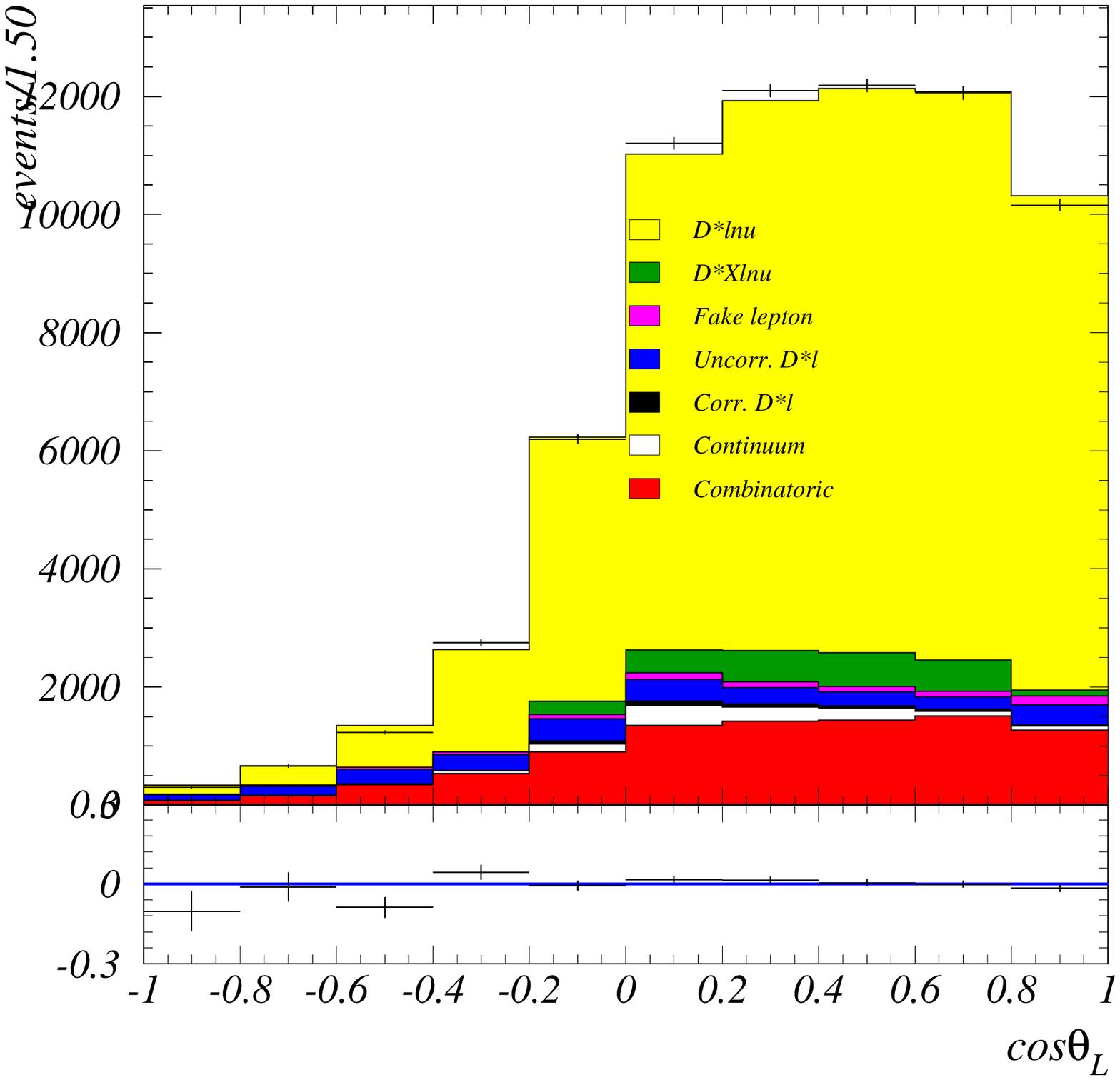}
\\
\includegraphics[width=0.50\textwidth,clip=]{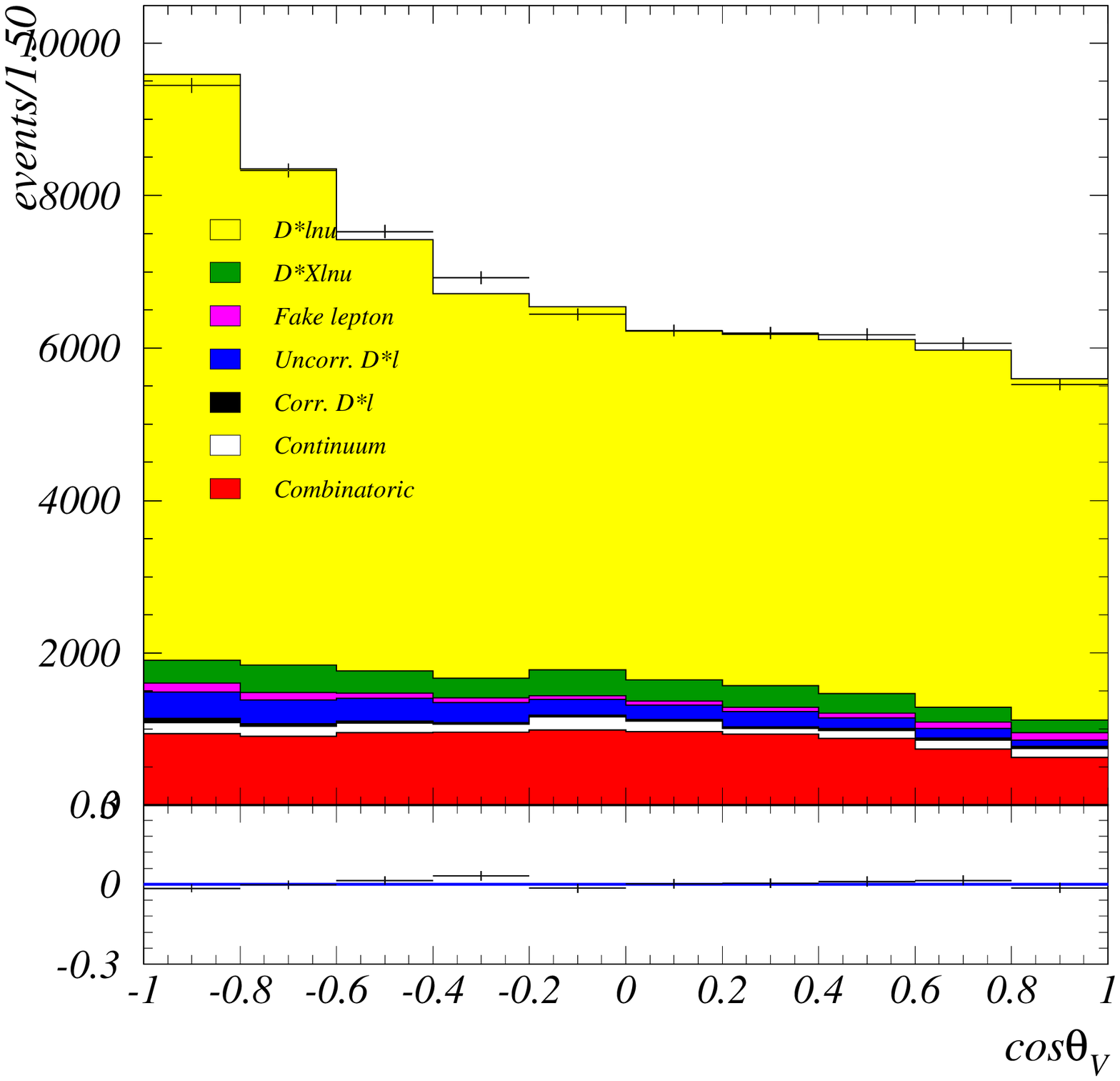}
&
\includegraphics[width=0.50\textwidth,clip=]{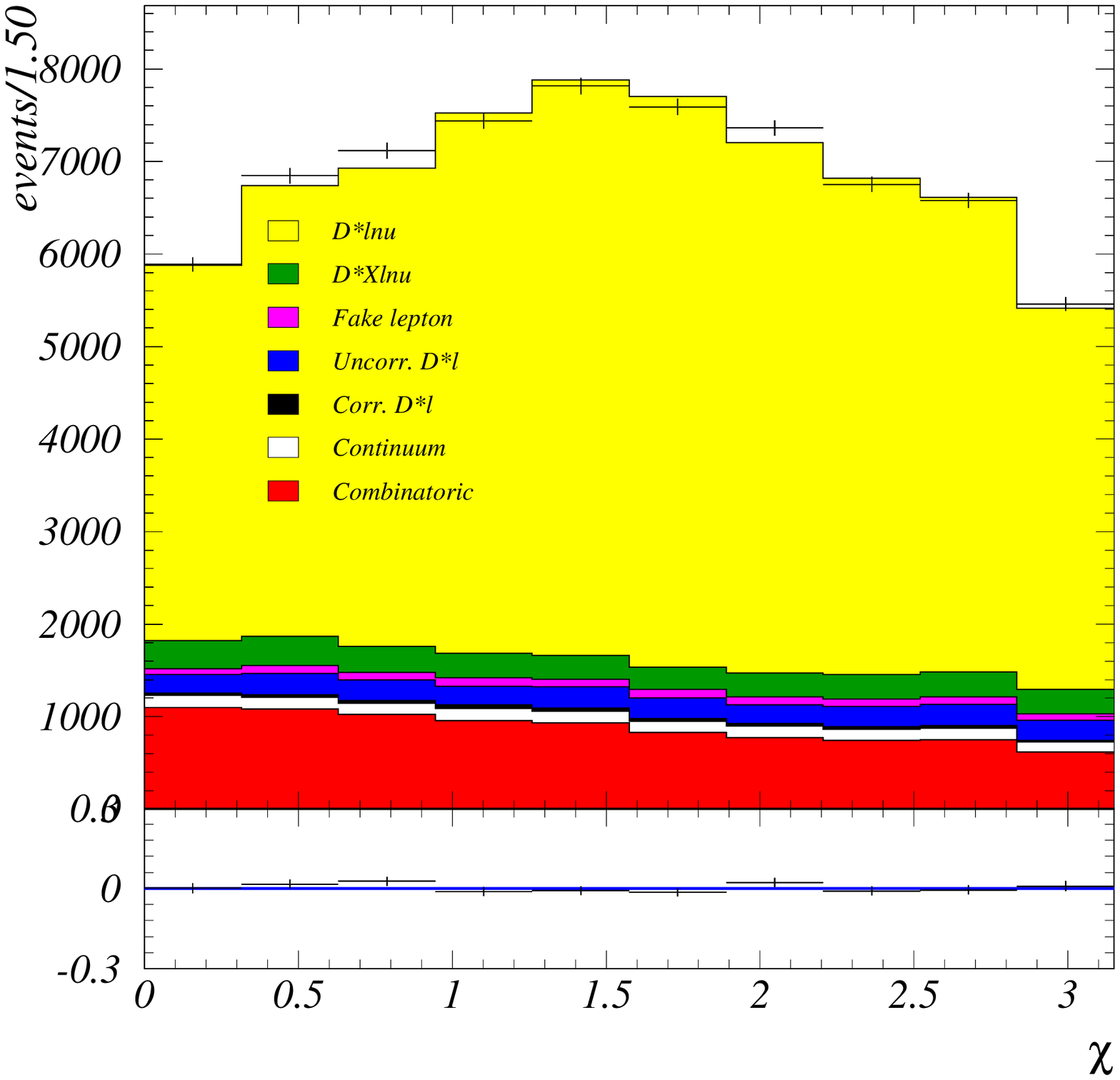}
\end{tabular}
\end{center}
\vspace*{-0.8cm}
\caption{Distributions of $w$, $\cos\theta_{\ell}$, $\cos\theta_{V}$,
  $\chi$ after the fit. The error bars represent the measured values,
  the colored histograms represent the background estimate and the fitted
  signal MC. The bottom plots show the ratio between the measurement and the
  sum of all background and fitted signal MC events. }
\label{fig:anchi_pnw}
\end{figure}

Figure~\ref{fig:anchi_pnw} shows the distributions of
the kinematical observables describing the \BztoDslnu decay with the
simulation prediction computed using
the best fit values for $|V_{cb}|$ and for the form factors.

The \BztoDslnu decay branching ratio is shown in the last row of the
Table~\ref{tab:stdresu}. The uncertainties are obtained through the
propagation of the ones on $\rho^2, R_1, R_2$ and
$\mathcal{F}(1)|V_{cb}|$, taking into account their correlations.

\begin{table}[htb]
\begin{center}
\begin{tabular}{|l||c|c|c|c|c|c|}
\hline
Parameter & Best value & $\sigma_{tot}$ & $\sigma_{stat}$ & $\sigma_{syst}$ \\
\hline
\hline
$\rho^2$& 1.156 & 0.095 & 0.094 & 0.004 \\
\hline 
$R_1$  & 1.329 & 0.131 & 0.131 & 0.007 \\
\hline
$R_2$  & 0.859 & 0.077 & 0.077 & 0.002 \\
\hline
$\mathcal{F}(1)|V_{cb}| \times 10^3$ & 35.03 & 0.95 & 0.39 & 0.86 \\
\hline \hline
BR(\BztoDslnu) \% & 4.83 & 0.24 & 0.05 & 0.24 \\
\hline
\end{tabular}
\end{center}
\caption{
Fit result for the standard analysis. $\sigma_{tot}$ is the
total uncertainty, and the last two columns show its
splitting in statistical and systematic contributions (from $\delta$
parameters). In the last row
the results for the branching ratio of \BztoDslnu obtained from the
fit results are shown.}
\label{tab:stdresu}
\end{table}

Table~\ref{tab:stdcorr} shows the correlation matrix for the standard
analysis result.

\begin{table}[htb]
\begin{center}
\begin{tabular}{|l||c|c|c|c|}
\hline
& $\rho^2$ & $R_1$ &  $R_2$ & $\mathcal{F}(1)|V_{cb}| \times 10^3$ \\
\hline \hline
$\rho^2$ &  $+1.000$ & $+0.864$ & $-0.922$ & $-0.028$ \\
\hline                                    
$R_1$ & $+0.864$ & $+1.000$ & $-0.928$ & $-0.226$ \\
\hline
$R_2$ & $-0.922$ & $-0.928$ & $+1.000$ & $+0.172$ \\
\hline                                    
$\mathcal{F}(1)|V_{cb}| \times 10^3$ & $-0.028$ & $-0.226$ & $+0.172$ & $+1.000$ \\
\hline 
\end{tabular}
\end{center}
\caption{Total correlation matrix for the standard analysis fit.}
\label{tab:stdcorr}
\end{table}

The goodness of fit can be evaluated directly from
the value of the fit function in the minimum, since it is based on a
least squares method, obtaining a $\chi^2$/d.o.f. = 23.99/24
corresponding to a $\chi^2$ probability of 46.2\%. In this evaluation the
number of degrees of freedom is considered equal to the number of used
bins, ignoring the bin correlations, minus the number of fitted
parameters. It is interesting also to check
the goodness of the fit on the separate distributions entering the fit
({\it i.e.}, $w$, \ctl and \ctv), considering for
each of them all ten bins in which their ranges have been
subdivided. The corresponding results are given in
Table~\ref{tab:partchi2}.

\begin{table}[htb]
\begin{center}
\begin{tabular}{|l|c|c|}
\hline
Variable & $\chi^2$/d.o.f. & $\chi^2$ probability \\
\hline                    
$w$ & 3.62/10 & 96.3\% \\
\hline
$\cos\theta_{\ell}$  & 12.93/10 & 22.8\% \\
\hline
$\cos\theta_{V}$ & 6.04/10 & 81.2\% \\
\hline                                    
\end{tabular}
\end{center}
\caption{Goodness of fit for the distributions entering the fit for
  the standard analysis. All ten bins are considered.}
\label{tab:partchi2}
\end{table}

\begin{figure}[htp]
\begin{center}
\begin{tabular}{cc}
\includegraphics[width=0.350\textwidth,clip=]{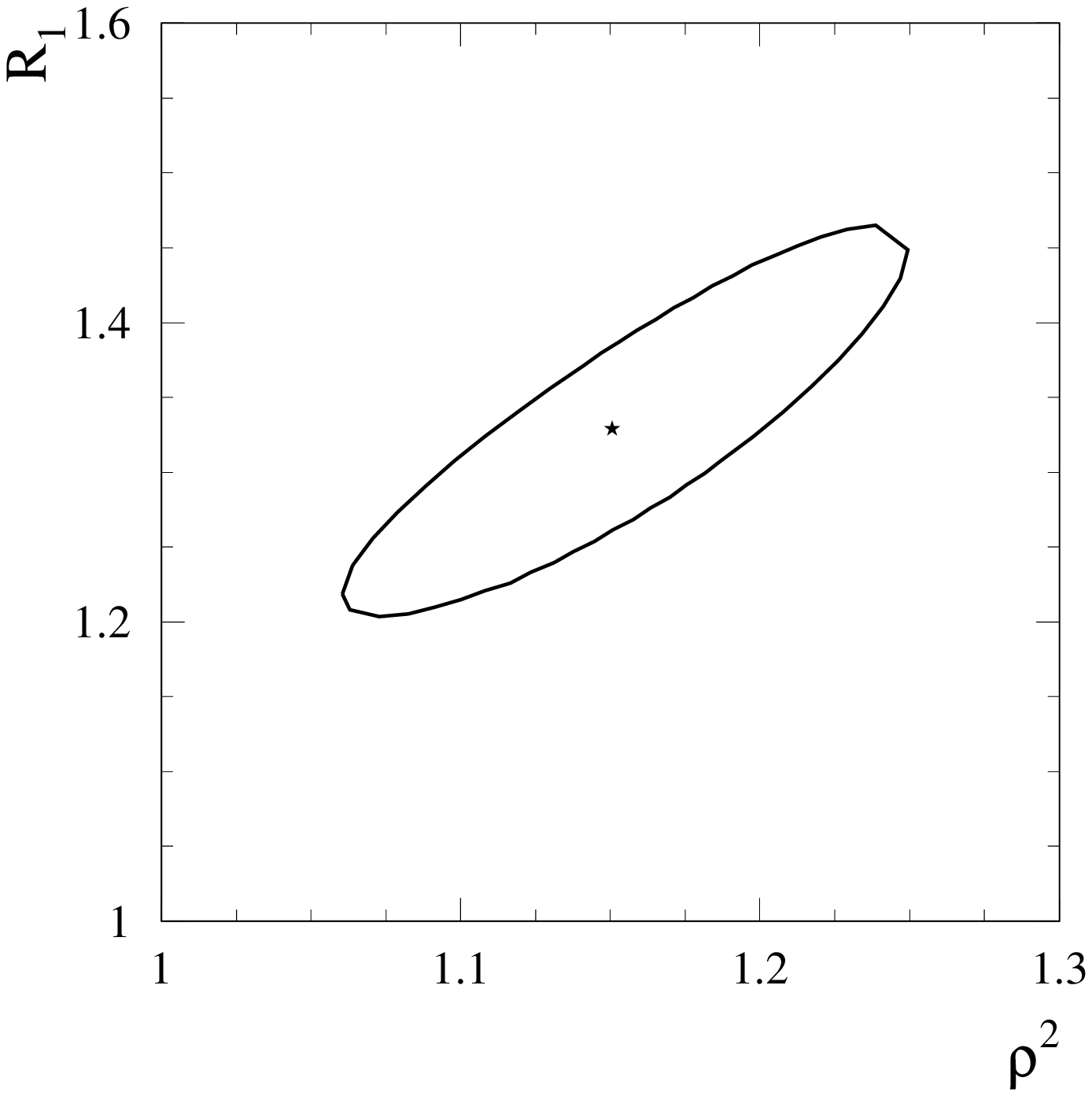}
&
\includegraphics[width=0.350\textwidth,clip=]{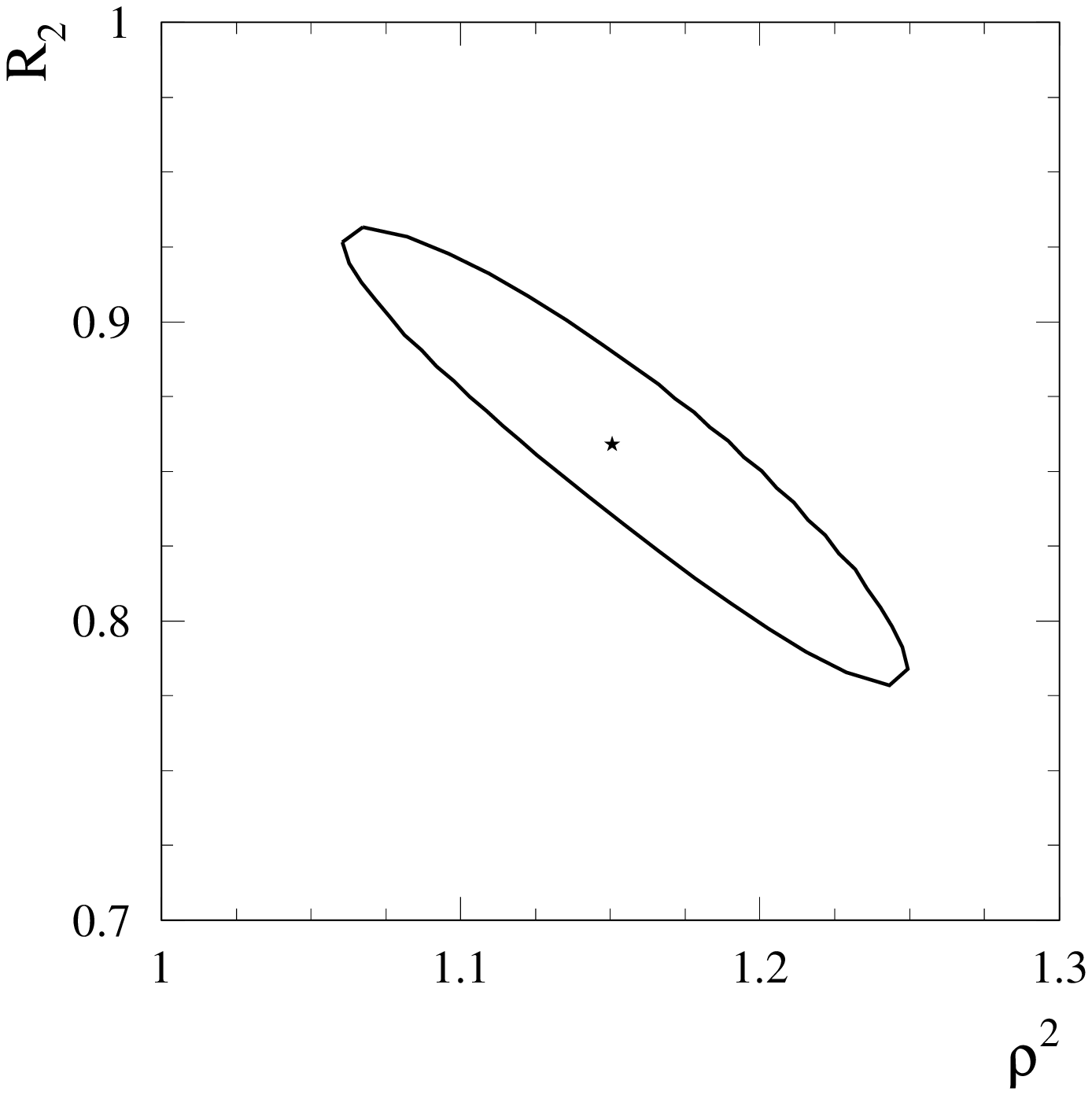}
\\
\includegraphics[width=0.350\textwidth,clip=]{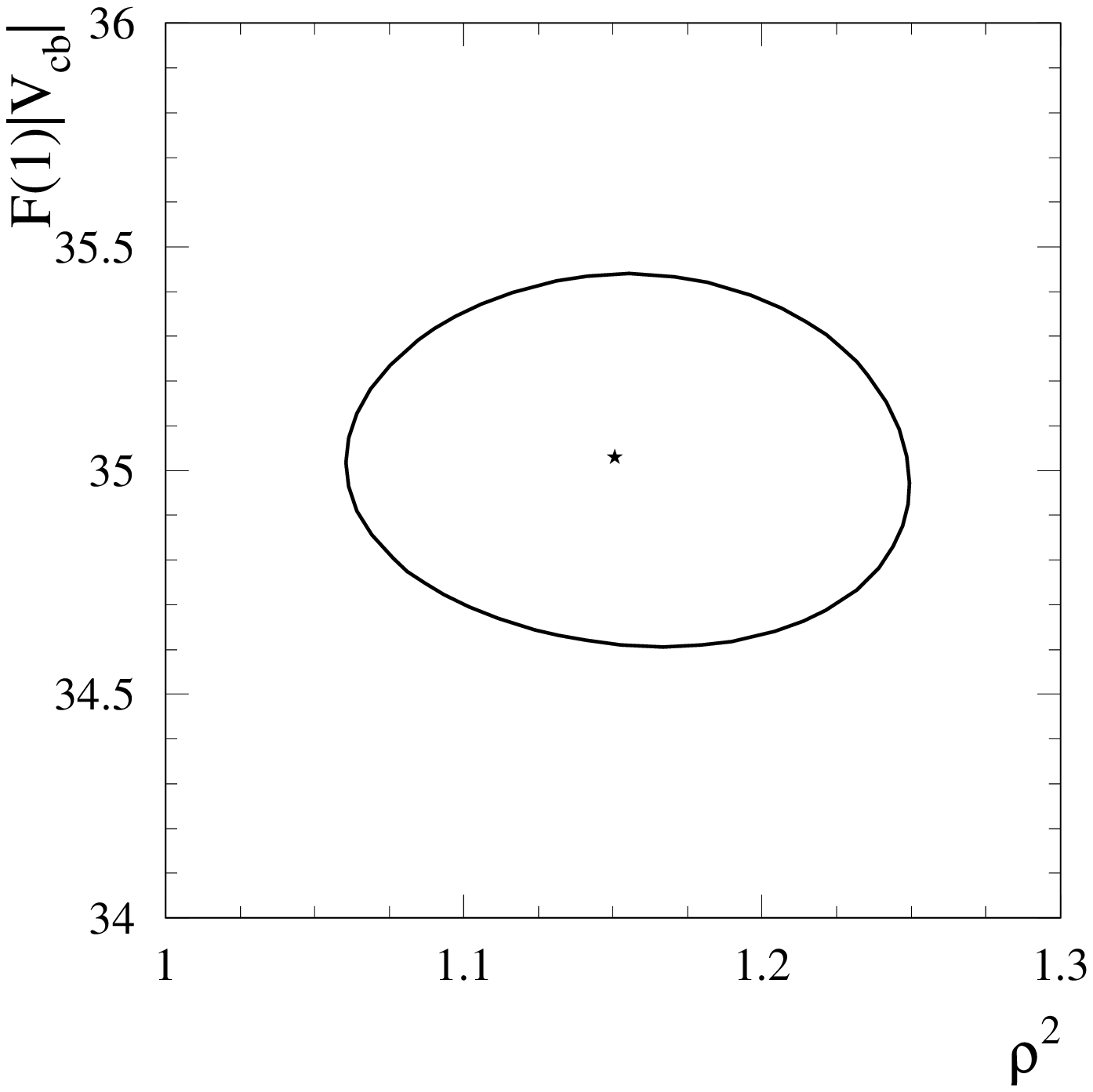}
&
\includegraphics[width=0.350\textwidth,clip=]{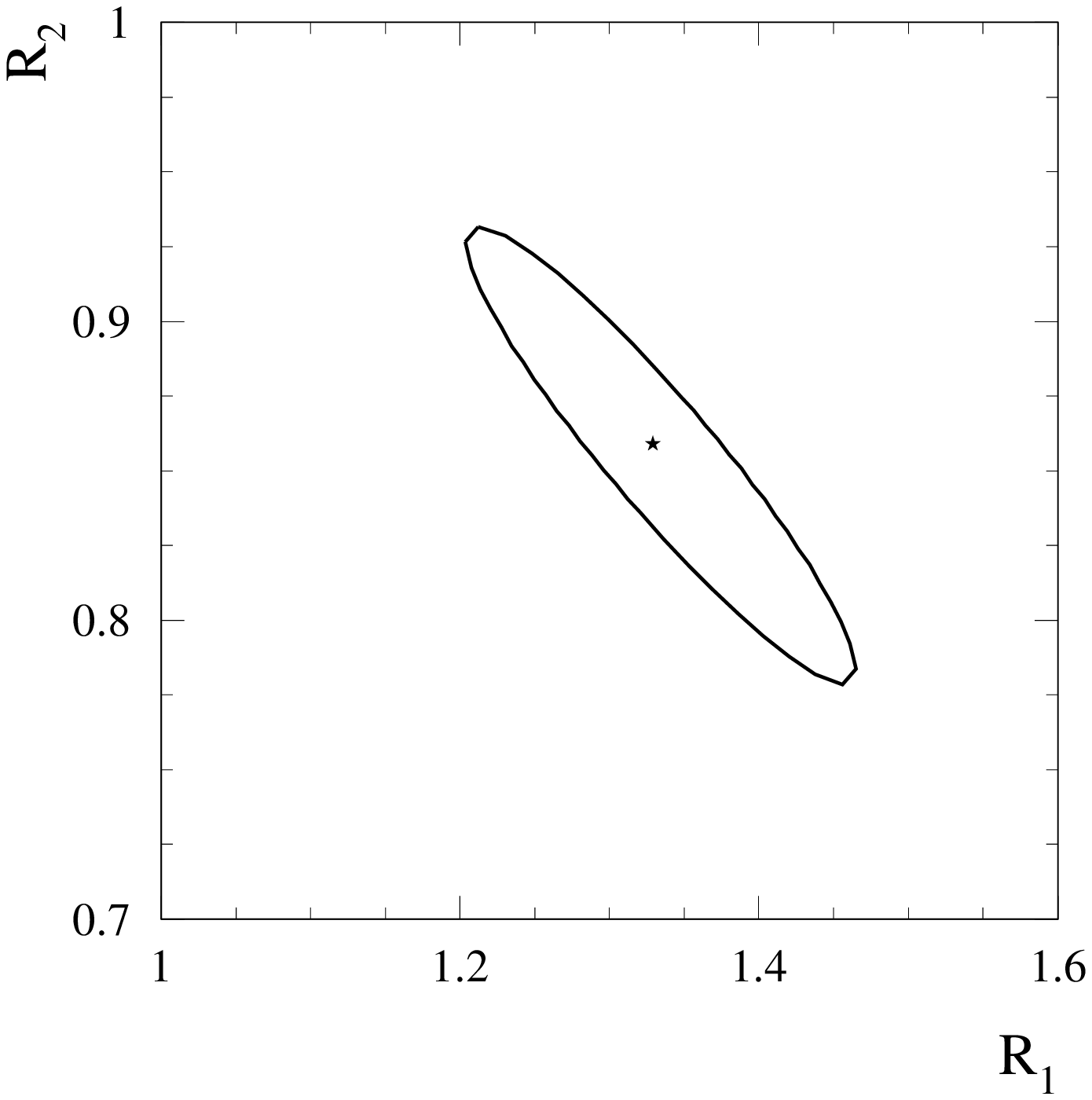}
\\
\includegraphics[width=0.350\textwidth,clip=]{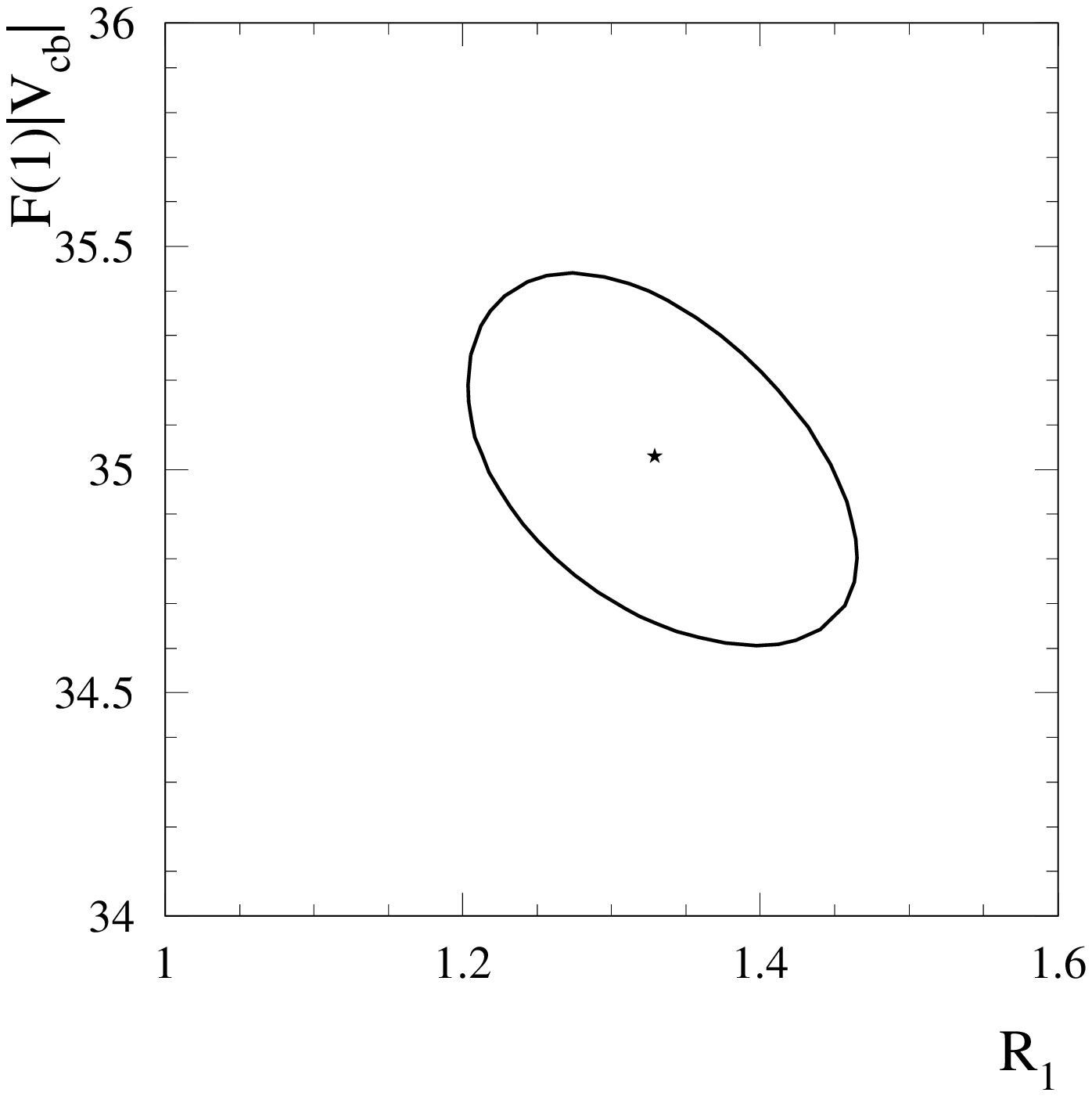}
&
\includegraphics[width=0.350\textwidth,clip=]{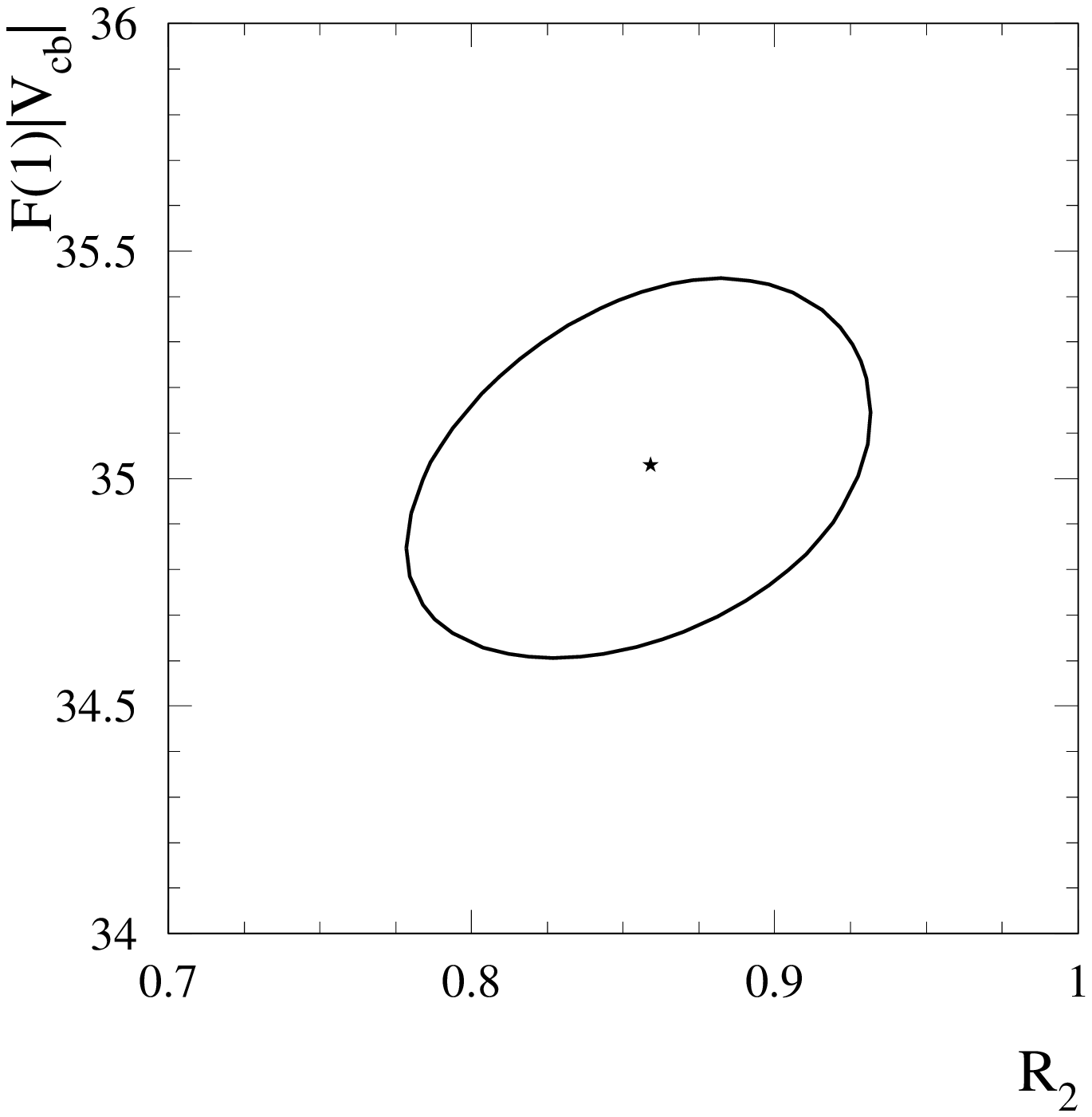}
\end{tabular}
\end{center}
\vspace*{-0.8cm}
\caption{
  Contour plot at 39\% C.L. ($\Delta \chi^2 = 1$) in the $\rho^2$-$R_1$, $\rho^2$-$R_2$, 
  $\rho^2$-$\mathcal{F}(1)|V_{cb}| \times 10^3$, $R_1$-$R_2$,  $R_1$-$\mathcal{F}(1)|V_{cb}| \times 10^3$,
  $R_2$-$\mathcal{F}(1)|V_{cb}| \times 10^3$ planes (keeping the
  $\delta$ parameters for the systematics at their best fit values).}
\label{fig:rho2r1_rho2r2_r1r2_rho2vcb_r1vcb_r2vcb_pnw}
\end{figure}

In Fig.~\ref{fig:rho2r1_rho2r2_r1r2_rho2vcb_r1vcb_r2vcb_pnw}
the contour plot at the 39\% C.L. ($\Delta \chi^2 = 1$) in the
$\rho^2$-$R_1$, $\rho^2$-$R_2$,  $R_1$-$R_2$,
$\rho^2$-$\mathcal{F}(1)|V_{cb}| \times 10^3$,
$R_1$-$\mathcal{F}(1)|V_{cb}| \times 10^3$,
$R_2$-$\mathcal{F}(1)|V_{cb}| \times 10^3$ planes are
shown, illustrating the correlations among the variables.

\subsection{Systematic uncertainties}

The summary of systematic uncertainties on the measured parameters is
presented in Table~\ref{table:totsys}. In the last column, the
uncertainty breakdown for the \BztoDslnu branching ratio is also
shown.

\begin{table*}[ht]
\begin{center}
\begin{tabular}{|l|c|c|c|c|c|}
\hline
CLN    & $\rho^2$ & $R_1$ & $R_2$ & $\mathcal{F}(1)|V_{cb}| \times 10^3$ &
BR(\BztoDslnu) \\
\hline
\hline
{\bf stat. error}
        &  {\bf 0.094}  &  {\bf 0.131}   & {\bf 0.077}    &{\bf 0.39}
        & {\bf 0.05} \\
\hline
\hline

PID, trk, BR $D^0$ & 0.004 & 0.007 & 0.002 & 0.86 & 0.24 \\
\hline

$\pi$ soft         & 0.013 & 0.005 & 0.001 & 0.46 & 0.18 \\
\hline

$D^*l$ vertex      & 0.014 & 0.010 & 0.008 & 0.06 & 0.06 \\
\hline

$B$ mom.           & 0.013 & 0.040 & 0.017 & 0.29 & 0.14 \\
\hline

rad. corr.         & 0.005 & 0.004 & 0.000 & 0.19 & 0.07 \\
\hline

$D^{**}$ compos.
                   & 0.011 & 0.008 & 0.009 & 0.10 & 0.07 \\
\hline

background cov. matrix    & 0.006 & 0.004 & 0.002 & 0.04 & 0.04 \\

\hline
\hline

{\bf partial sum}
               &{\bf 0.027}  &  {\bf 0.043}   & {\bf 0.021}    &{\bf
               1.04} & {\bf 0.35} \\
\hline
\hline

$B^0$ lifetime    & - & - & - & 0.10 & 0.03\\
\hline

$B\bar{B}$ number & - & - & - & 0.19 & 0.05 \\  
\hline
BR($D^* \rightarrow D^0\pi$) & - & - & - & 0.13 & 0.04 \\
\hline
$f_{++}/f_{00}$ & 0.008 & 0.010 & 0.006 & 0.41 & 0.15 \\
\hline
\hline

{\bf tot. syst. error}
               &{\bf 0.028}   &  {\bf 0.044}  & {\bf 0.022}    &{\bf 1.15} &{\bf 0.39} \\
\hline

\hline
\end{tabular}
\end{center}
\caption{
Breakdown of systematics uncertainties.}
\label{table:totsys}
\end{table*}

\subsubsection{Statistical, PID and tracking efficiency and {\boldmath $D^0$}
  branching fraction uncertainties}

The uncertainty on the parameters given by the fit is not of purely
statistical origin, since the systematic uncertainty sources which are
not common to all events (PID efficiency, tracking efficiency, and
$D^0$ branching fraction uncertainties) are taken directly into account
in the fit through the $\delta$ coefficients appearing in the weights
$\mbox{W}_i^{S,k}$. The splitting between statistical and systematic
part of the fit result uncertainty is obtained by computing the purely
statistical covariance matrix (by fixing all $\delta$ parameters
at their best value) and subtracting it from the global one.

\subsubsection{Soft pion efficiency}

The uncertainty coming from the soft pion efficiency correction is
determined by propagating the uncertainties of the parameters of the
efficiency correction function. In addition, an overall efficiency
uncertainty due to track reconstruction and selection is considered (a
scale factor affecting only $\mathcal{F}(1)|V_{cb}|$), its magnitude
being 1.3\%.  The two uncertainties are added in quadrature.

\subsubsection{{\boldmath $D^*l$} vertex reconstruction efficiency}

The uncertainty coming from the vertex reconstruction method used to
select the $D^*l$ system has been evaluated using two alternative
choices to the standard method: the first corresponds to removing the
beam spot constraint from the vertex fit, while the second corresponds
to removing the lepton information from the vertex fit. The biggest
variation is taken as the systematic uncertainty.

\subsubsection{{\boldmath $B$} momentum}

A correction to the $B$ momentum in the simulation has to be applied
to obtain agreement with data. Effectively this implies rescaling the
\cBDl value determined for the simulation by a 0.97 factor. Half of the
relative effect of the correction is considered as systematic
uncertainty on it.

\subsubsection{Radiative corrections}

Radiative corrections to the \BztoDslnu decay are computed in the
simulation by {\tt PHOTOS}~\cite{ref:photos}, which describes the
final state photon radiation (FSR) up to $\mathcal{O}(\alpha^2)$.  In
the event reconstruction no attempt to recover photons emitted in the
decay is performed, therefore the simulated prediction of the
reconstructed event is sensitive to the detail of the radiative
corrections.  This is particularly important in the
$\cos\TBY$ fit for electrons, where the FSR is responsible
for the long tail at low values.

No detailed calculation of the full $\mathcal{O}(\alpha)$ radiative
corrections to the \BztoDslnu decay is available in literature.
Recently, a new $\mathcal{O}(\alpha)$ calculation of radiative
corrections in kaon decays has become available~\cite{ref:andre04}
explicitly taking into account radiated photons, and a new detailed
comparison with {\tt PHOTOS} has been presented. From this comparison
it is evident that the radiated photon's energy spectrum is quite well
reproduced by {\tt PHOTOS}, the main difference being in the angular
distribution of the photons with respect to the lepton.

In order to assess the systematic uncertainty due to the imperfect
treatment of the radiative corrections, we have used the comparison
presented in Ref.~\cite{ref:andre04} to reweight our simulated events in
order to reproduce the photon angular spectrum given there, for
photons above 10 MeV in the $B$ center of mass. The effect of the
reweighting has been used as an estimate of the systematic
uncertainty. In this approach the possible effect of the $B^0$ decay
form factors on the FSR description is completely ignored.

\subsubsection{{\boldmath $B \rightarrow \Dstar \ellp
  \nul X$} background description}

The shapes of the different components of the $B \rightarrow \Dstar \ellp
  \nul X$ background are
taken from the simulation. In order to account for the uncertainty on
their knowledge, the fit has been performed using only one of them at
a time. The study has been performed for the \DzKP subsample only, and
assumed to be valid for all subsamples. The uncertainty on the
result is given by half of the biggest variation observed.

\subsubsection{Fit covariance matrix}

The effect of using the background shape evaluated in different
projections to compute the background covariance matrix is used as an
evaluation of the corresponding systematic uncertainty (the background
overall normalization is fixed to the average of the one computed in
the projections).  The maximal variation with respect to the basic
result when changing the projection used is taken as an estimate.

\subsubsection{Global normalization factors}

In the second part of Table~\ref{table:totsys} the effect of the
uncertainty on quantities which are normalization factors is shown.
The effect of the uncertainties in the $D^0$ branching ratios has already been
discussed, and it affects all parameters because it is
changing the relative amount of signal events coming from different
$D^0$ decays. On the other hand, the $B^0$ lifetime, the $D^* \rightarrow
D^0\pi$ branching ratio, and the $B\bar{B}$ measured number are global
scale factors affecting only $\mathcal{F}(1)|V_{cb}|$.  The
uncertainty on the ratio $f_{++}/f_{00} = \mathcal{B}(\Upsilon(4S)
\rightarrow B^+B^-)/\mathcal{B}(\Upsilon(4S) \rightarrow B^0\bar{B}^0)$
affects both the absolute number of measured \BztoDslnu decays (and
therefore $\mathcal{F}(1)|V_{cb}|$) and the relative ratio of
background events from $B^0$ and $B^{\pm}$ in the simulation. This
second aspect influences the $\cos\TBY$ shapes and therefore
the background determination. For this reason this uncertainty source
cannot be considered a pure scale uncertainty affecting only
$\mathcal{F}(1)|V_{cb}|$, but it touches also the form factors
parameters. The analysis has been repeated by changing the
$f_{++}/f_{00}$ value by one sigma and the variation has been used as
systematic uncertainty on the results from this source.
The $B^0$ lifetime used is $\tau_{B^0} = 1.532 \pm
0.009$~\cite{ref:pdg04}. The $D^{*} \rightarrow D^0 \pi$ branching
ratio used is $67.7 \pm 0.5 \%$. The $f_{++}/f_{00}$ value used is
$1.055 \pm 0.055$ from Ref.~\cite{ref:pdg02}.
The systematic uncertainty on the number of $B\bar{B}$ ($N_{B\bar{B}}$) is
1.1\%.  The uncertainties on these input quantities are propagated to
the final fit results.

\section{CONCLUSIONS}
\label{sec:summary}

\subsection{Summary of results of this analysis}

A sample of about 52800 fully reconstructed \BztoDslnu decays
collected by the \babar\ detector has been used to measure 
both $\mathcal{F}(1)|V_{cb}|$ and the form factor parameters in the
CLN parameterization~\cite{ref:CLNpaper} $\rho^2$, $R_1$, and
$R_2$ in a global fit. Both identified electrons and muons have been used, the \Dstarm
candidates are reconstructed from the \Dstarm\ra\Dzb\pim\ decay and
the \Dzb candidate is reconstructed in three different decay modes
$K^-\pi^+$, $K^-\pi^+\pi^+\pi^-$ and $K^-\pi^+\pi^0$.

The obtained results are (where the first errors are statistical and the second
systematic):
\begin{eqnarray*}
  \mathcal{F}(1)|V_{cb}| & = & (35.03 \pm 0.39 \pm 1.15) \times 10^{-3} \\
  \rho^2 & = & 1.156 \pm 0.094 \pm 0.028 
  \\
  R_1 & = & 1.329 \pm 0.131 \pm 0.044 \\
  R_2 & = & 0.859 \pm 0.077 \pm 0.022 
\end{eqnarray*}

The correlations between the fitted parameters are:
\begin{eqnarray*}
  \rho(\rho^2,R_1) & = & +86\% \\
  \rho(\rho^2,R_2) & = & -92\% \\
  \rho(\rho^2,\mathcal{F}(1)|V_{cb}|) & = & -3\% \\
  \rho(R_1,R_2) & = & -92\% \\
  \rho(R_1,\mathcal{F}(1)|V_{cb}|) & = & -23\% \\
  \rho(R_2,\mathcal{F}(1)|V_{cb}|) & = & +17\%
\end{eqnarray*}

Using a recent lattice calculation~\cite{ref:auno} ($h_{A_{1}}(1) =
\mathcal{F}(1) = 0.919^{+ 0.030}_{- 0.035}$) results in the following
value for $|V_{cb}|$:
\begin{equation*}
  |V_{cb}| = ( 38.12 \pm 0.42 \pm 1.25 ^{+ 1.24}_{- 1.45} ) \times
   10^{-3},
\end{equation*}
where the third error is due to the uncertainty in $\mathcal{F}(1)$.

The corresponding branching fraction for the decay \BztoDslnu is
found to be:
\begin{equation*}
  \mathcal{B}(\BztoDslnu) = (4.84 \pm 0.05 \pm 0.39 ) \%.
\end{equation*}

\subsection{Combination of results with alternate form factor parameters 
measurement}
The \babar\ collaboration recently published a measurement~\cite{ref:bad1224} 
of the same form factor parameters in \BztoDslnu
decays based on an unbinned maximum likelihood fit to the 
four-dimensional decay distribution using a subset of the data 
analyzed in this work.  This earlier analysis has a higher sensitivity to 
the form factor parameters and resulted in 
$\rho^2 = 1.145 \pm 0.066 \pm 0.035$,
$R_1 =  1.396 \pm 0.070 \pm 0.027, $ and $
R_2 =  0.885 \pm 0.046 \pm 0.013$.

We can combine the two form factor parameters measurements taking into 
account the correlation between them and obtain

\begin{eqnarray*}
\mathcal{F}(1)|V_{cb}| & = & ( 34.68 \pm 0.32 \pm 1.15) \times 10^{-3}  \\
\rho^2 & = & 1.179 \pm 0.048 \pm 0.028 \\
R_1 & = & 1.417 \pm 0.061 \pm 0.044 \\
R_2 & = & 0.836 \pm 0.037 \pm 0.022.
\end{eqnarray*}
The statistical errors are significantly improved compared to the 
analysis presented in this paper.
The two analyses have largely the same sources of systematic 
uncertainties, and thus we retain  the systematic measurement 
errors established in this paper.  The combined statistical errors 
are still larger than the systematic errors, but not by a large 
factor.
 
The correlation coefficients between the fitted parameters are:
\begin{eqnarray*}
\rho(\rho^2,R_1) & = & +70\% \\
\rho(\rho^2,R_2) & = & -83\% \\
\rho(\rho^2,\mathcal{F}(1)|V_{cb}|) & = & +27\% \\
\rho(R_1,R_2) & = & -84\% \\
\rho(R_1,\mathcal{F}(1)|V_{cb}|) & = & -39\% \\ 
\rho(R_2,\mathcal{F}(1)|V_{cb}|) & = & +22\% 
\end{eqnarray*}

Using the lattice calculation for $\mathcal{F}(1)$, we obtain an improved 
value for $|V_{cb}|$~\cite{ref:auno}:
$$|V_{cb}| =(37.74 \pm 0.35 \pm  1.25 \pm ^{1.23}_{1.44} ) \times 10^{-3},$$ 
\noindent
where the third error reflects the current uncertainty on  $\mathcal{F}(1)$.  
This combined measurement of $|V_{cb}|$ represents a significant 
improvement over previous measurements.  It supersedes all previous 
exclusive \babar\ measurements based on partial data sets or 
less detailed analyses.

The corresponding branching fraction of the decay \BztoDslnu is
found to be:
\begin{equation*}
  \mathcal{B}(\BztoDslnu) = (4.77 \pm 0.04 \pm 0.39 ) \%.
\end{equation*}

\par
\section{ACKNOWLEDGMENTS}
\label{sec:Acknowledgments}
 
\input acknowledgements

%% file: acknowledgements.tex
We are grateful for the 
extraordinary contributions of our \pep2\ colleagues in
achieving the excellent luminosity and machine conditions
that have made this work possible.
The success of this project also relies critically on the 
expertise and dedication of the computing organizations that 
support \babar.
The collaborating institutions wish to thank 
SLAC for its support and the kind hospitality extended to them. 
This work is supported by the
US Department of Energy
and National Science Foundation, the
Natural Sciences and Engineering Research Council (Canada),
Institute of High Energy Physics (China), the
Commissariat \`a l'Energie Atomique and
Institut National de Physique Nucl\'eaire et de Physique des Particules
(France), the
Bundesministerium f\"ur Bildung und Forschung and
Deutsche Forschungsgemeinschaft
(Germany), the
Istituto Nazionale di Fisica Nucleare (Italy),
the Foundation for Fundamental Research on Matter (The Netherlands),
the Research Council of Norway, the
Ministry of Science and Technology of the Russian Federation, and the
Particle Physics and Astronomy Research Council (United Kingdom). 
Individuals have received support from 
the Marie-Curie IEF program (European Union) and
the A. P. Sloan Foundation.

%% file: paper.bbl
\begin{thebibliography}{99}
\bibitem{footnote1} Charge conjugate decay modes are implicitly included, $\ell =e,\mu$.
\end{thebibliography}
